\UseRawInputEncoding
\documentclass[superscriptaddress, twocolumn, amsmath, amssymb, aps,pre, notitlepage,longbibliography]{revtex4-1} 
\usepackage{graphicx,graphics,epsfig,subfigure,times,bm,bbm,amssymb,amsmath,amsfonts,amsthm,mathrsfs,MnSymbol}
\usepackage[matrix,frame,arrow]{xypic}
\usepackage[normalem]{ulem}
\usepackage{slashed}
\usepackage{dcolumn}
\usepackage{tabularx}
\usepackage{color}
\usepackage[usenames,dvipsnames,svgnames,table]{xcolor}
\usepackage[english]{babel}
\usepackage{verbatim}
\definecolor{darkblue}{rgb}{0.0,0.0,0.3}
\usepackage[colorlinks=true,
            linkcolor=red,
            urlcolor= darkblue,
            citecolor=blue]{hyperref}

\newcommand{\bea}{\begin{eqnarray}}
\newcommand{\eea}{\end{eqnarray}}

\begin{document}
\title{Quantum Carnot thermal machines re-examined: Definition of efficiency and the effects of strong coupling}

\author{Junjie Liu}
\email{jjliu.fd@gmail.com}
\address{Department of Physics, International Center of Quantum and Molecular Structures, Shanghai University, Shanghai, 200444, China}
\author{Kenneth A. Jung}
\email{kj6821@gmail.com}
\address{275 Hawthorne Avenue, Palo Alto, California, 94301, USA}

\begin{abstract}
Whether the strong coupling to thermal baths can improve the performance of quantum thermal machines remains an open issue under active debate. Here, we revisit quantum thermal machines operating with the quasi-static Carnot cycle and aim to unveil the role of strong coupling in maximum efficiency. Our analysis builds upon definitions of excess work and heat derived from an exact formulation of the first law of thermodynamics for the working substance, which captures the non-Gibbsian thermal equilibrium state that emerges at strong couplings during quasi-static isothermal processes. These excess definitions differ from conventional ones by an energetic cost for maintaining the non-Gibbsian characteristics. With this distinction, we point out that one can introduce two different yet thermodynamically allowed definitions for efficiency of both the heat engine and refrigerator modes. We dub them excess and hybrid definitions which differ in the way of defining the gain for the thermal machines at strong couplings by either just analyzing the energetics of the working substance or instead evaluating the performance from an external system upon which the thermal machine acts, respectively. We analytically demonstrate that the excess definition predicts that the Carnot limit remains the upper bound for both operation modes at strong couplings, whereas the hybrid one reveals that strong coupling can suppress the maximum efficiency rendering the Carnot limit unattainable. These seemingly incompatible predictions thus indicate that it is imperative to first gauge the definition for efficiency before elucidating the exact role of strong coupling, thereby shedding light on the on-going investigation on strong-coupling quantum thermal machines.  
\end{abstract}

\date{\today}

\maketitle

\section{Introduction}
Quantum thermal machines (QTMs) \cite{Benenti.17.PR,Mitchison.19.CP,Myers.22.AVSQS} perform energetic tasks such as energy conversion and extraction at the nanoscale. With quantum systems as working substances, QTMs are capable of harnessing quantum resources to facilitate the operation process and achieve unparalleled capabilities that are provably impossible with classical counterparts \cite{Buffoni.19.PRL,Carrega.22.PRXQ,JiW.22.PRL,Lu.23.PRB}. As such, QTMs are becoming potential platforms for demonstrating the quantumness of energetic tasks and contrasting classical and quantum thermodynamics, as highlighted by recent intriguing theoretical proposals \cite{Buffoni.19.PRL,Carrega.22.PRXQ,Lu.23.PRB,Quan.07.PRE,Brandner.15.PRX,Proesmans.16.PRX,Newman.17.PRE,Watanabe.17.PRL,Perarnau-Llobet.18.PRL,Mohammady.19.PRE,Ono.20.PRL,Strasberg.21.PRL,Strasberg.21.PRE,Liu.21.PRL} and delicate experimental realizations \cite{JiW.22.PRL,Abah.12.PRL,Rossnagel.14.PRL,Robnagel.16.S,Sood.16.NP,Peterson.19.PRL,Klatzow.19.PRL,Assis.19.PRL,Pekola.19.ARCMP,Lindenfels.19.PRL,Kim.22.NP,BuJ.23.PRL,Koch.23.N} to name just a few. 

Concerning the thermodynamics of QTMs, it is recognized that the description of QTMs does not permit a complete adoption of the well-established classical framework as QTMs can display distinct features beyond the capability of the classical description. One feature that attracts recent attention asserts that working substances of QTMs can experience possible strong system-bath couplings as their surface area can become comparable to their volume \cite{Abah.12.PRL,Robnagel.16.S,Peterson.19.PRL,Klatzow.19.PRL,Lindenfels.19.PRL,Ono.20.PRL,JiW.22.PRL,Josefsson.18.NN}, rendering the weak-coupling assumption inapplicable. Investigating strong-coupling QTMs becomes urgent. On the one hand, the field of strong-coupling quantum thermodynamics has witnessed significant progresses over the decades with a number of self-consistent strategies formulated (see a recent review \cite{Talkner.20.RMP} and references therein). Nevertheless, no consensus on a universal framework for strong-coupling quantum thermodynamics has been reached as one lacks prior knowledge of thermodynamic behavior of systems at strong couplings. Applying the existing strong-coupling quantum thermodynamic frameworks to QTMs, one can reveal their own thermodynamic signatures of strong coupling that allow for verification with current experimental capabilities, thereby providing benchmarks for establishing a universal framework for strong-coupling quantum thermodynamics. On the other hand, strong coupling can enable non-negligible system-bath correlation and entanglement which are useful operational resources for QTMs, in this regard, a recent study \cite{Longstaff.23.PRA} suggests that the strong coupling is inevitable when devising steady state entanglement QTMs. 

To date, substantial efforts have been put into the investigation of strong-coupling QTMs \cite{Gallego.14.NJP,Gardas.15.PRE,Kato.16.JCP,Strasberg.16.NJP,Newman.17.PRE,Xu.18.PRE,Restrepo.18.NJP,Perarnau-Llobet.18.PRL,Newman.20.PRE,Liu.21.PRL,Shirai.21.PRR,Carrega.22.PRXQ,Ivander.22.PRE,Liu.22.PRE,Latune.23.PRA,Kaneyasu.23.PRE}. To circumvent the theoretical and numerical challenges imposed by strong couplings, one typically focuses on either specific models and methodologies \cite{Carrega.22.PRXQ,Newman.17.PRE,Gallego.14.NJP,Kato.16.JCP,Strasberg.16.NJP,Restrepo.18.NJP,Ivander.22.PRE,Liu.22.PRE} or specific thermodynamic settings \cite{Perarnau-Llobet.18.PRL,Liu.21.PRL,Newman.20.PRE,Shirai.21.PRR,Latune.23.PRA,Kaneyasu.23.PRE}, yielding somewhat contradictory conclusions on the role of strong couplings. Whereas some studies claim that strong coupling could bring up operational advantages to potentially enhance the performance of QTMs \cite{Strasberg.16.NJP,Xu.18.PRE,Shirai.21.PRR,Carrega.22.PRXQ,Liu.21.PRL,Longstaff.23.PRA,Kaneyasu.23.PRE,Latune.23.PRA}, there are opposite perspectives indicating that strong coupling is merely detrimental to the operation of QTMs \cite{Gallego.14.NJP,Kato.16.JCP,Newman.17.PRE,Restrepo.18.NJP,Perarnau-Llobet.18.PRL,Newman.20.PRE,Ivander.22.PRE,Liu.22.PRE}. Thus, elucidating the exact role of strong coupling in the performance of QTMs still warrants further investigation.

%
\begin{figure*}[thb!]
 \centering
\includegraphics[width=1.9\columnwidth]{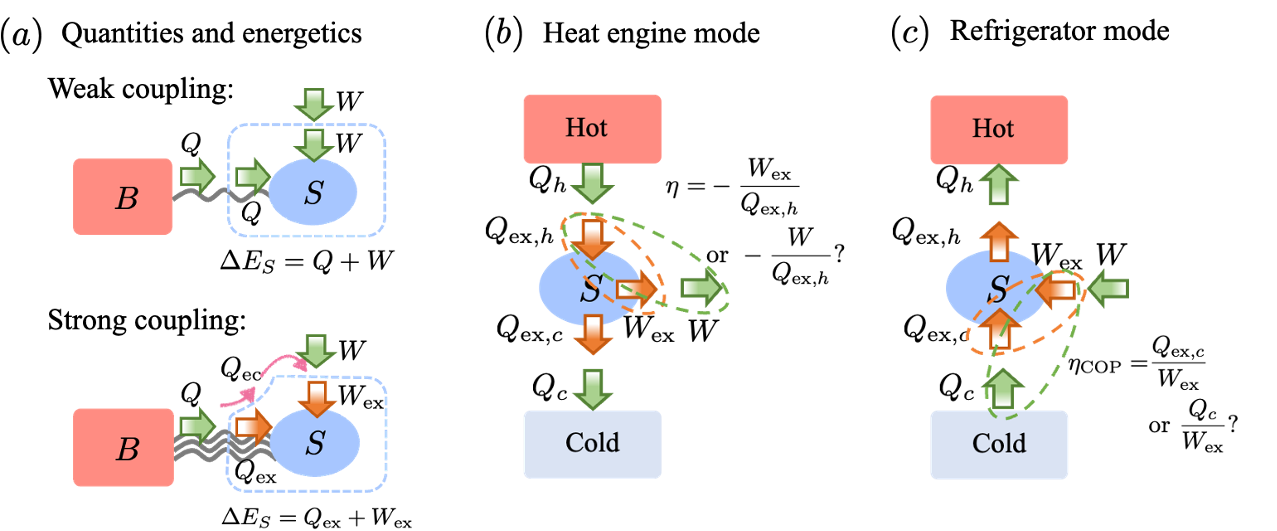} 
\caption{Sketch of the study. (a) Quantities and energetics during a quasi-static isothermal process. Upper panel: Weak coupling limit with conventional heat definition $Q$ due to bath energy loss, and conventional work definition $W$ due to external drivings which are not shown in the figure. The first law of thermodynamics for the system reads $\Delta E_S=Q+W$ [see Eq. (\ref{eq:3})]. Lower panel: Strong coupling scenario with both conventional quantities $Q$, $W$ and excess quantities $Q_{\mathrm{ex}}=Q-Q_{\mathrm{ec}}$ [see Eq. (\ref{eq:ds})], $W_{\mathrm{ex}}=W+Q_{\mathrm{ec}}$ [see Eq. (\ref{eq:new_fl})] with $Q_{\mathrm{ec}}$ an energetic cost for maintaining the non-Gibbsian thermal state. In this scenario, the first law of thermodynamic for the system reads $\Delta E_S=Q_{\mathrm{ex}}+W_{\mathrm{ex}}$ [Eq. (\ref{eq:new_fl})].  Directions of arrows reflect our positive sign convention of thermodynamic quantities. (b) Energy flow representation of a heat engine mode at strong couplings. Due to the difference between $W$ and $W_{\mathrm{ex}}$, one may come up with different definitions for thermodynamic efficiency $\eta$ which differ in gain as highlighted by the orange and green dashed circles (see details in Sec. \ref{sec:3}). (c) Energy flow representation of a refrigerator mode at strong couplings. Due to the difference between $Q_c$ and $Q_{\mathrm{ex},c}$, one may carry out different definitions for the coefficient of performance $\eta_{\mathrm{COP}}$ which differ in gain as highlighted by the orange and green dashed circles (see details in Sec. \ref{sec:4}).
}
\protect\label{fig:first_law}
\end{figure*}
Here we focus on QTMs operating with a quasi-static Carnot cycle and analyze the strong coupling effect on their maximum efficiency. Noting that the thermal equilibrium state can deviate from a canonical Gibbsian form at strong couplings \cite{Cresser.21.PRL,Anto-Sztrikacs.23.PRXQ}, we utilize the excess heat and work definitions proposed in Ref. \cite{Gardas.15.PRE} which naturally account for such a scenario through an exact formulation of the first law of thermodynamics for the working substances at strong couplings. Remarkably, these excess definitions ensure thermodynamic relations between quantities to satisfy the same forms across the weak to strong couplings in the quasi-static limit, thereby providing a promising strategy to lift the ambiguity in defining heat and work at strong couplings. Compared with conventional heat and work definitions based on the change of system density matrix and Hamiltonian, respectively, the excess counterparts differ by an energetic cost term indispensable for maintaining the non-Gibbsian characteristics.

We note that excess work and heat sufficiently determine the energetics of the working substances, whereas the conventional ones, when applying at strong coupling, are instead associated with external work and heat sources (see Fig. \ref{fig:first_law} (a) for an illustration). With this distinction, we point out that there is a flexibility in defining a thermodynamic figure of merit for QTMs at strong couplings, depending on how one defines the gain of applying the QTMs in that limit. For clarity, here we just consider the efficiency (coefficient of performance, COP) as the corresponding figure of merit for the heat engine (refrigerator) mode.

Specifically, we propose that one can define two thermodynamically-allowed definitions for the chosen figure of merit at strong couplings. The first definition involves just the excess quantities in an analogy with the weak coupling scenario where one utilizes quantities in the first law of thermodynamics to define the figure of merit. We dub it an ``excess" definition for later convenience. The excess definition treats the excess work (excess heat from the cold bath) as the gain of applying a heat engine (refrigerator), thereby evaluating the performance from solely the system energetics in a way similar to the weak coupling scenario; See the orange dashed circles in Fig. \ref{fig:first_law} (b) and (c). Alternatively, one can recall the design purpose of QTMs and evaluate their performance from the standpoint of the external systems upon which the QTMs act. In this spirit, one should take the conventionally-defined work (conventionally-defined heat out of the cold bath) as the gain for a heat engine (refrigerator), thereby yielding a ``hybrid" definition which involves both excess and conventionally-defined quantities; See green dashed circles in Fig. \ref{fig:first_law} (b) and (c).

These two definitions for the figure of merit coincide only at weak couplings where the energetic cost for maintaining the non-Gibbsian feature vanishes and both reduce to the Carnot limit as expected in the quasi-static limit. However, a distinction between the two definitions emerges at strong couplings. We demonstrate that the excess definition still recovers the Carnot limit for both quasi-static quantum heat engine and quantum refrigerator modes at strong couplings, whereas thehybrid definition leads one to a conclusion that strong coupling plays a negative role as it suppresses the maximum efficiency, rendering the Carnot limit unattainable at strong couplings. Therefore, our results indicate that an uncertainty in the definition for the figure of merit would hinder a thorough evaluation of the strong coupling effect on the optimal performance. Nevertheless, focusing on the quasi-static limit, we suggest that the hybrid definition is more preferable than the excess one in practical applications as it takes into account the overall effect of QTMs exerted on external systems and thus closely reflects the designed purpose of QTMs. 

This paper is organized as follows. For the sake of completeness, we first recap excess heat and work definitions \cite{Gardas.15.PRE} in Sec. \ref{sec:1} by considering an arbitrary quasi-static isothermal process during which the system can strongly couple to the thermal bath. In Sec. \ref{sec:2}, we present details of the adopted quantum Carnot cycle and expressions for excess quantities using relations carried out in Sec. \ref{sec:1}. Then in Sec. \ref{sec:3} and \ref{sec:4}, we analyze the heat engine mode and refrigerator mode, respectively. We explicitly propose two different yet thermodynamically-allowed definitions for the figure of merit of both operation modes. We further demonstrate the strong coupling effect on optimal performance using both definitions. Finally, we present few remarks and conclude the study in Sec. \ref{sec:5}.

\section{Excess heat and work at strong couplings}\label{sec:1}
To analyze thermodynamic performance of QTMs, one should first settle down the definitions for heat and work. At weak couplings, one usually starts from the changes of system energy and identify the part associated with an entropic change as the heat \cite{Weimer.08.EPL,Alipour.22.PRA} and attribute the remaining part to the work. However, definitions for heat and work are subtle at strong couplings and no consensus on their definitions has been reached yet \cite{Talkner.20.RMP}. Here we adopt the framework developed in Ref. \cite{Gardas.15.PRE} which guarantees the applicability of the aforementioned thermodynamic interpretations of heat and work \cite{Weimer.08.EPL,Alipour.22.PRA} at strong couplings. For the sake of completeness and to fix the notation, we will briefly recap the definitions for the excess quantities defined in Ref. \cite{Gardas.15.PRE} with a particular emphasis on clarifying their physical meanings which will be the basis for proposing and constrasting different definitions for the figure of merit at strong couplings.

Considering a {\it quasi-static} isothermal process in which a quantum system coupled to a single thermal bath at a temperature $T=\beta^{-1}$ (setting $\hbar=1$ and $k_B=1$ hereafter) can stay in an instantaneous thermal equilibrium state. When the system-bath coupling strength deviates from the weak coupling regime, the system equilibrium state $\rho_S$ is no longer of a Gibbsian form $\rho_{th}=e^{-\beta H_S}/Z_S$ with the system Hamiltonian $H_S$ and $Z_S=\mathrm{Tr}[e^{-\beta H_S}]$ \cite{Cresser.21.PRL,Anto-Sztrikacs.23.PRXQ},
\begin{equation}
\rho_S~=~\mathrm{Tr}_B\left[\frac{e^{-\beta H_{\mathrm{tot}}}}{Z_{\mathrm{tot}}}\right]~\neq~\rho_{th}.
\end{equation}
Here, $H_{\mathrm{tot}}$ denotes the total Hamiltonian including the system, the bath ($B$) and the system-bath interaction, $Z_{\mathrm{tot}}=\mathrm{Tr}_{\mathrm{tot}}[e^{-\beta H_{\mathrm{tot}}}]$ and $\mathrm{Tr}_B$  denotes a trace operation over bath degrees of freedom. 

We consider the von Neumann entropy of the system state $S=-\mathrm{Tr}[\rho_S\ln\rho_S]$ which can become a thermodynamic entropy when one has perfect knowledge of the system \cite{Strasberg.21.PRXQ}. The system von Neumann entropy $S$ satisfies \cite{Parrondo.15.NP,Gardas.15.PRE}
\begin{equation}\label{eq:first_law}
S ~=~ \beta (E_S-\mathcal{F}).
\end{equation}
Here, $E_S=\mathrm{Tr}[H_S\rho_S]$ denotes the internal energy of the system, $\mathcal{F}=F+TD(\rho_S||\rho_{th})$ is the nonequilibrium free energy \cite{Deffner.12.A,Deffner.13.PRX,Parrondo.15.NP} with $F=-T \ln Z_S$ the Helmholtz free energy and $D(\rho_S||\rho_{th})=\mathrm{Tr}[\rho_S(\ln\rho_S-\ln\rho_{th})]$ the quantum relative entropy between $\rho_S$ and $\rho_{th}$ accounting for the non-Gibbsian nature of $\rho_S$ at strong couplings. We remark that in the quantum domain the strong-coupling equilibrium state $\rho_S$ can contain nonzero coherence in the energy basis of the system Hamiltonian \cite{Anto-Sztrikacs.23.PRXQ}, therefore the quantum relative entropy involved in the nonequilibrium free energy definition is not equivalent to the Kullback–Leibler divergence between classical distributions. Hence, the definitions presented below cannot be directly applied to classical systems. Taking $S$ as the thermodynamic entropy \cite{Strasberg.21.PRXQ}, we can then regard $\mathcal{F}$ as a thermodynamic free energy at strong couplings. It is evident that $\mathcal{F}\ge F$ with the equality attained at weak couplings where $\rho_S=\rho_{th}$, implying an additional capacity to perform work using information of relative entropy \cite{Deffner.13.PRX} at strong couplings. We remark that Eq. (\ref{eq:first_law}) defines an exact form of the first law of thermodynamics for isothermal processes under an arbitrary coupling strength.

To define heat and work applicable at strong couplings, one can start from considering infinitesimal changes of thermodynamic quantities at hand. We first have
\bea\label{eq:3}
\delta E_S &=& \mathrm{Tr}[\delta \rho_S H_S]+\mathrm{Tr}[\rho_S\delta H_S]\nonumber\\
&\equiv& \delta Q+\delta W.
\eea
Here, $\delta W\equiv \mathrm{Tr}[\rho_S\delta H_S]$ represents the work generated by external driving fields in the system Hamiltonian, $\delta Q\equiv\mathrm{Tr}[\delta \rho_S H_S]$ represents the heat exchanged with the thermal bath \cite{Gardas.15.PRE}. To see this, we denote the total Hamiltonian $H_{\mathrm{tot}}=H_B+V+H_S$ with $H_B$ and $V$ the bath Hamiltonian and system-bath interaction, respectively. We first have $\delta W=\delta \mathrm{Tr}[H_{\mathrm{tot}}\rho]=\mathrm{Tr}[\rho\delta H_S]$ with $\rho$ the total density matrix by noting $\mathrm{Tr}[H_{\mathrm{tot}}\delta\rho]=0$ as a result of the quantum Liouville equation for $\rho$. We then find $\delta \mathrm{Tr}[H_{\mathrm{tot}}\rho]=\delta \mathrm{Tr}[H_B\rho]+\delta \mathrm{Tr}[H_S\rho]+\delta \mathrm{Tr}[V\rho]\simeq \delta \mathrm{Tr}[H_B\rho_B]+\delta E_S=\delta W$ with $\delta \mathrm{Tr}[V\rho]$ being negligible for a constant coupling in the quasi-static limit and $\rho_B$ the reduced bath state. Comparing it with Eq. (\ref{eq:3}), we can find $\delta Q=-\delta\mathrm{Tr}[H_B\rho_B]$. One recalls that the usual semi-classical modelings of cyclic QTMs do not explicitly include pure work sources and loads in the Hamiltonian but assume their presence through external driving fields in the system Hamiltonian \cite{Esposito.10.NJP}. In this sense, we emphasize that $W$ defines the amount of work that can be harnessed by external work loads even at strong couplings. Hereafter, we set the sign convention that positive heat and work increase the internal energy of the system. 

At weak couplings, $Q$ and $W$ unambiguously define the heat and work, respectively, and quantify the energetics of the working substance according to Eq. (\ref{eq:3}) [see the upper panel of Fig. \ref{fig:first_law} (a)]. In a quasi-static isothermal process, we have $\rho_S=\rho_{th}$ diagonal in the energy basis, hence the heat $Q$ induces a pure entropic change. However, this is no longer the case at strong couplings as we will show below. Nevertheless, we remark that one can still use $Q$ and $W$ at strong couplings as they are still well-defined thermodynamic quantities in that regime [see the lower panel of Fig. \ref{fig:first_law} (a)].

Inserting Eq. (\ref{eq:3}) into Eq. (\ref{eq:first_law}), we find
\begin{equation}\label{eq:ds}
\delta S~=~\beta(\delta Q-\delta Q_{\mathrm{ec}})~\equiv~\beta \delta Q_{\mathrm{ex}}.
\end{equation}
Here, $\delta Q_{\mathrm{ec}}\equiv \delta \mathcal{F}-\delta W$ denotes an energetic cost for maintaining the non-Gibbsian characteristics \cite{Gardas.15.PRE} and corresponds to an energy part of $\delta Q$ that does not induce an entropic change. $\delta Q_{\mathrm{ex}}\equiv \delta Q-\delta Q_{\mathrm{ec}}$ thus defines an excess heat accompanying with a pure entropic change, in accordance with the definition for a thermodynamic heat \cite{Weimer.08.EPL,Alipour.22.PRA}. Hence, we take $Q_{\mathrm{ex}}$ as the appropriate heat definition for the working substance at strong couplings in the quasi-static limit. Notably, $Q_{\mathrm{ex}}$ reduces to $Q$ at weak couplings as $Q_{\mathrm{ec}}$ vanishes in that limit.

With the definition of excess heat, we can rewrite Eq. (\ref{eq:3}) as
\begin{equation}\label{eq:new_fl}
\delta E_S~=~\delta W_{\mathrm{ex}}+\delta Q_{\mathrm{ex}}.
\end{equation}
Here, the excess work $\delta W_{\mathrm{ex}}\equiv \delta W+\delta Q_{\mathrm{ec}}=\delta \mathcal{F}$, reducing to $\delta W$ at weak couplings, precisely corresponds to the nonequilibrium free energy change, in accordance with the conventional expectation for the work in the quasi-static limit. 

From the first law of thermodynamics, Eq. (\ref{eq:first_law}), one generally finds $\Delta \mathcal{F}(t)=\Delta E_S(t)-T\Delta S(t)$ with $\Delta A(t)=A(t)-A(0)$ for an arbitrary quantity $A$. Meanwhile, we have $\Delta E_S(t)=Q_{\mathrm{ex}}(t)+W_{\mathrm{ex}}(t)$ in terms of excess quantities which, together with the second law $T\Delta S(t)\ge Q_{\mathrm{ex}}(t)$, implies $\Delta E_S(t)~\le~T\Delta S(t)+W_{\mathrm{ex}}(t)$. Altogether, we arrive at a principle of maximum work applicable at strong couplings $-\Delta \mathcal{F}(t)~\ge~-W_{\mathrm{ex}}(t)$ with the equality taken in the quasi-static limit. Hence, one can regard $-W_{\mathrm{ex}}$ as the maximum work that the working substance can provide during a quasi-static isothermal process according to our sign convention. As such, we take $ W_{\mathrm{ex}}$ as the work definition for the working substance at strong couplings which together with the excess heat $Q_{\mathrm{ex}}$ completely determines the energetics of the working substance (see the lower panel of Fig. \ref{fig:first_law} (a) for an illustration). However, as we remarked before, only the part $ W$ of $W_{\mathrm{ex}}$ can be harnessed by external systems. Noting $Q_{\mathrm{ec}}=W_{\mathrm{ex}}-W$, one can also interpret $Q_{\mathrm{ec}}$ as the amount of work that cannot be harnessed by external systems.

\section{Quantum Carnot cycle}\label{sec:2}
To operate QTMs, we consider a {\it quasi-static} Carnot cycle consisting of two quasi-static isothermal strokes and two quasi-static adiabatic strokes. Taking the heat engine mode as an example, we have four strokes following the given order: 

(i) A quasi-static hot isothermal expansion stroke. During this stroke, a quantum working substance is attaching to a hot thermal bath at temperature $T_h$ with the system Hamiltonian changing from $H_S^A$ to $H_S^B$; We use superscripts to distinguish quantities at different stages induced by, e.g., tuning control parameters. To realize expansion of quantum working substances with discrete energy levels \{$E_n$\}, one needs to change all energy levels $E_n^A\to E_n^B=\xi E_n^A$ with the same ratio $0<\xi<1$ \cite{Quan.07.PRE}.  Taking a two-level system for example, the expansion amounts to decreasing the energy gap. During this stroke, we can use definitions in the previous section to express excess work and heat in terms of the nonequilibrium free energy and entropy changes, respectively,
\bea
W_{\mathrm{ex},h} &=& \mathcal{F}^B-\mathcal{F}^A,\nonumber\\
Q_{\mathrm{ex},h} &=& T_h(S^B-S^A).
\eea
We also have a formal definition $W_h\equiv\int_A^B \mathrm{Tr}[\rho_S(t)\frac{d}{dt}H_S(t)]dt$ which denotes the work generated by driving fields during this stroke. In practice, once $H_S(t)$ and the governing equation of motion for $\rho_S(t)$ are specified during an isothermal process, we can estimate the nonequilibrium free energy and entropy changes directly, thereby obtaining the values of the excess work and heat, respectively. $W_h$ can also be estimated using its formal definition which, together with the relation $Q_{\mathrm{ec},h}=W_{\mathrm{ex},h}-W_h$ and the value of $W_{\mathrm{ex},h}$, further gives an evaluation of the energetic cost.

(ii) A quasi-static adiabatic expansion stroke. During this stroke, the working substance is detached from the hot thermal bath, rendering heat exchanges impossible leaving the system entropy unchanged, $S^C=S^B$. The temperature of the working substance drops from $T_h$ to $T_c$ accomplished by an internal energy change which comes in the form of work. By requiring that all energy gaps of the working substance change by the same ratio $T_c/T_h$, one can ensure that the end points of the adiabatic expansion stroke coincide with those of an isothermal processes \cite{Quan.07.PRE}. Hence we can still adopt the first law of thermodynamics depicted in the previous section to rewrite the work,
\bea
W_{\mathrm{ex},a1} &=& W_{a1}~=~E_S^C-E_S^B\nonumber\\
&=& \mathcal{F}^C-\mathcal{F}^B-(T_h-T_c)S^B.
\eea

(iii) A quasi-static cold isothermal compression stroke. During this stroke, the working substance is attaching to a cold thermal bath at temperature $T_c$ with all energy levels increasing $E_n^C\to E_n^D=\xi' E_n^C$ by the same ratio $\xi'>1$. For a two-level system, the compression is thus realized by increasing the energy gap. We have
\bea
W_{\mathrm{ex},c} &=& \mathcal{F}^D-\mathcal{F}^C,\nonumber\\
Q_{\mathrm{ex},c} &=& T_c[S^D-S^B].
\eea
And similarly, $W_c\equiv\int_C^D \mathrm{Tr}[\rho_S(t)\frac{d}{dt}H_S(t)]dt$.

(iv) A quasi-static adiabatic compression stroke. During this stroke, the system is detached from the cold thermal bath, and the temperature of the working substance increases from $T_c$ to $T_h$ along with all energy gaps of the working substance increasing by the same ratio $T_h/T_c$. Similar to the second adiabatic stroke, we find
\bea
W_{\mathrm{ex},a2} &=& W_{a2}~=~E_S^A-E_S^D\nonumber\\
&=& \mathcal{F}^A-\mathcal{F}^D+(T_h-T_c)S^A.
\eea
Here, we have used the fact that $S^D=S^A$. After this stroke, we couple the working substance to the hot thermal bath again, thus completing the thermodynamic cycle. The Carnot cycle for the refrigerator mode can be obtained by reversing the aforementioned orderings of strokes as well as each stroke's operational direction, and the associated thermodynamic quantities can be expressed in an analogous way.

In the following section we will distinguish between two possible sets of definitions for the figure of merit of quantum Carnot thermal machines (see Fig. \ref{fig:first_law} (b) and (c)). The first definition which we dub an {\it excess} one involves just excess quantities, amounting to characterizing the performance from solely the system energetics in complete analogy with the weak coupling scenario. The second definition instead evaluates the performance from the action on an external system and inevitably contains both the conventional and excess quantities, thereby bearing the name of a {\it hybrid} definition.

\section{Quantum Carnot heat engine}\label{sec:3}
We first focus on the cyclic quantum Carnot heat engine operating with the Carnot cycle stated in Sec. \ref{sec:2}. At weak couplings, one utilizes $Q_h$  and total work $W\equiv\sum_{i=h,a1,c,a2}W_i$ to define the efficiency which, however, cannot be directly applied to the strong coupling regime \cite{Gardas.15.PRE,Liu.21.PRL}. Hence, to unveil the strong coupling effect on the efficiency, one needs to decipher an appropriate definition for the efficiency first. In this regard, one should bear in mind that a thermodynamic efficiency quantifies how worth a gain at the expense of a cost is.  

For a heat engine operated at strong couplings, the actual cost should be the excess heat $Q_{\mathrm{ex}}^h$ during the hot isothermal process that the working substance absorbs. As for the gain, we point out that one can have two possible choices at strong couplings due to the distinction between $W_{\mathrm{ex}}\equiv\sum_{i=h,a1,c,a2}W_{\mathrm{ex},i}$ and $W$.

Through introducing excess heat and work that have the same interpretations as their weak-coupling counterparts, one actually builds up complete analog between weak and strong couplings in terms of the first law of thermodynamics [Eq. (3) for weak coupling and Eq. (5) for strong coupling extended to a two-bath scenario]. At weak couplings, one can just utilize $W$ and $Q_h$ involved in the first law of thermodynamics to define a thermodynamic efficiency which recovers the Carnot bound in the quasi-static limit. Following this spirit and noting the analog, one can define a thermodynamic efficiency using just excess quantities at strong couplings
\begin{equation}\label{eq:in_eta}
\eta^{\mathrm{ex}}~\equiv~-\frac{W_{\mathrm{ex}}}{Q_{\mathrm{ex},h}}.
\end{equation}
This definition amounts to evaluating the performance of heat engines from solely the energetics of the working substance and treating $W_{\mathrm{ex}}$ as the gain in analogy with the weak coupling scenario. We dub it an {\it excess} definition. Using results in Sec. \ref{sec:2}, one identifies
\begin{equation}\label{eq:sumWex}
W_{\mathrm{ex}}~=~(T_h-T_c)[S^A-S^B].
\end{equation}
Inserting Eq. (\ref{eq:sumWex}) into Eq. (\ref{eq:in_eta}), one finds that the inside efficiency definition exactly coincides with the Carnot efficiency 
\begin{equation}\label{eq:eta_in}
\eta^{\mathrm{ex}}~=~\frac{T_h-T_c}{T_h}=\eta_c.
\end{equation}
Hence, one concludes that the strong couplings have no impact on the maximum efficiency when defining the efficiency of a heat engine from a complete analogy with the weak coupling scenario. This result was first obtained in Ref. \cite{Gardas.15.PRE} and was stated as a manifestation of the thermodynamic universality of the Carnot efficiency.

Alternatively, as we remarked before, only the part $W$ can be directly harnessed by external systems as it is associated with external driving fields. Hence, if one evaluates the performance of a quantum Carnot heat engine from the standpoint of external systems, one then naturally refers to $W$ as the gain of the machine since a part of the excess work manifests as an energetic cost which cannot be harnessed by external systems. This viewpoint brings up a thermodynamic efficiency definition
\begin{equation}\label{eq:out_eta}
\eta^{\mathrm{hyb}}~\equiv~-\frac{W}{Q_{\mathrm{ex},h}}~=~\eta_c+\frac{\sum_{i=h,c}Q_{\mathrm{ec},i}}{Q_{\mathrm{ex},h}}.
\end{equation}
In arriving at the second equality, we have used the relation $W=W_{\mathrm{ex}}-\sum_{i=h,c}Q_{\mathrm{ec},i}$ and Eq. (\ref{eq:eta_in}). As this efficiency definition involves both conventional and excess quantities, we refer to it as a {\it hybrid} definition. It is evident that $\eta^{\mathrm{hyb}}$ and $\eta^{\mathrm{ex}}$ are identical only at weak couplings where $Q_{\mathrm{ec},i}$ vanishes. At strong couplings, the two definitions $\eta^{\mathrm{ex}}$ and $\eta^{\mathrm{hyb}}$ are no longer equivalent. We note that the second term on the right-hand-side of Eq. (\ref{eq:out_eta}) arises when $Q_{\mathrm{ec},i}$ becomes nonzero. Recalling that $Q_{\mathrm{ec},i}$ quantifies the energetic cost for maintaining the non-Gibbsian state during the isothermal strokes, we can attribute the second term to the effect of strong coupling.

To ensure that the hybrid efficiency $\eta^{\mathrm{hyb}}$ is thermodynamically allowed, we need to check whether it can surpass the Carnot efficiency due to the presence of an extra term. To this end, we can analyze the sign of this extra term which is solely determined by the numerator $\sum_{i=h,c}Q_{\mathrm{ec},i}$ as the denominator $Q_{\mathrm{ex},h}$ is always positive in a heat engine. We first note that $-W_{\mathrm{ex}}>0$ and $-W>0$ in a heat engine according to our sign convention. With the principle of maximum work, we know that $-W_{\mathrm{ex}}$ corresponds to the maximum work that the engine can generate as it is just the nonequilbrium free energy change during a cycle, while $-W$ is the amount of the work that can be harnessed by an external work load eventually. It is reasonable to expect that the external work load cannot gain more work than the engine can provide, hence one should have $|W_{\mathrm{ex}}|\ge|W|$ or equivalently, $-W_{\mathrm{ex}}\ge -W$. From this inequality, we can infer that $-\sum_{i=h,c}Q_{\mathrm{ec},i}=W-W_{\mathrm{ex}}\ge 0$ with the equality attained at weak couplings. Hence the second term on the right-hand-side of Eq. (\ref{eq:out_eta}) is always non-positive, leading to a constraint on the hybrid efficiency definition $\eta^{\mathrm{hyb}}$,
\begin{equation}
    \eta^{\mathrm{hyb}}~\le~\eta_c.
\end{equation}
Here, the equality is met at weak couplings where $\sum_iQ_{\mathrm{ec},i}=0$. At strong couplings with $\sum_iQ_{\mathrm{ec},i}<0$, one generally expects $\eta^{\mathrm{hyb}}<\eta_c$, rendering $\eta^{\mathrm{hyb}}$ a valid thermodynamic efficiency consistent with the second law of thermodynamics. More importantly, adopting $\eta^{\mathrm{hyb}}$ as the efficiency definition would lead one to conclude that the strong coupling has a negative effect in the sense that it prevents the maximum efficiency of the strong-coupling heat engine from reaching the Carnot efficiency, unlike their weak-coupling counterparts.

\section{Quantum Carnot refrigerator}\label{sec:4}
We now turn to a cyclic quantum Carnot refrigerator operating with a reversed Carnot cycle. The corresponding expressions for the excess work and heat are obtained by reversing the sign of those listed in Sec. \ref{sec:2}. 

To define a coefficient of performance (COP) for the refrigerator, one similarly needs to specify the cost and the resulting gain. Here, the cost should be the excess work $W_{\mathrm{ex}}$ which directly induces the free energy change of the working substance in the quasi-static limit. As for the gain, one also has two different choices, similar to the heat engine mode.

Here one still has an analogy between the weak and strong coupling scenarios after introducing excess thermodynamic quantities. Inspired by the COP definition $Q_c/W$ at weak couplings using just quantities involved in the first law of thermodynamics, one can also evaluate the performance of the refrigerator mode using just the excess quantities describing the energetics of the working substance at strong coupling,
\begin{equation}
    \eta_{\mathrm{COP}}^{\mathrm{ex}}~\equiv~\frac{Q_{\mathrm{ex},c}}{W_{\mathrm{ex}}}.
\end{equation}
Here, the excess heat $Q_{\mathrm{ex},c}$ during the cold isothermal stroke is identified to be the gain of the refrigerator mode by noting the analogy with the weak coupling scenario. We dub it an {\it excess} COP definition. For the reversed Carnot cycle, one has $Q_{\mathrm{ex},c}=T_c[S^B-S^A]$ and $W_{\mathrm{ex}}=(T_h-T_c)[S^B-S^A]$ which are opposite in sign to those shown in Sec. \ref{sec:2}, yielding
\begin{equation}\label{eq:cop_in}
    \eta_{\mathrm{COP}}^{\mathrm{ex}}~=~\eta_{\mathrm{COP}}^{c}.
\end{equation}
Here, $\eta_{\mathrm{COP}}^c\equiv T_c/(T_h-T_c)$ is just the classical Carnot limit of the COP. Namely, the optimal COP of the refrigerator at strong couplings can reach the Carnot limit just as was the case for the weak coupling scenario if adopting the excess definition for the COP, thereby implying no strong coupling effect on the optimal performance.

Nevertheless, during the cold isothermal expansion stroke, one recognizes that the total amount of energy extracted out of the cold bath should be $Q_c$ instead of $Q_{\mathrm{ex},c}$ that induces a system entropic change. Recall that the design purpose of a quantum absorption refrigerator is to extract energy out of a cold bath, one should instead treat $Q_c$ as the gain of the refrigerator. Therefore, one can have an alternative COP definition
\begin{equation}\label{eq:cop_out}
\eta_{\mathrm{COP}}^{\mathrm{hyb}}~\equiv~\frac{Q_c}{W_{\mathrm{ex}}}~=~\eta_{\mathrm{COP}}^{c}+\frac{Q_{\mathrm{ec},c}}{W_{\mathrm{ex}}}.
\end{equation}
In arriving at the second equality, we have used Eq. (\ref{eq:cop_in}) and the relation $Q_c=Q_{\mathrm{ex},c}+Q_{\mathrm{ec},c}$. Similar to the heat engine mode, we refer to $\eta_{\mathrm{COP}}^{\mathrm{hyb}}$ as a {\it hybrid} COP definition. We note that $\eta_{\mathrm{COP}}^{\mathrm{ex}}$ and $\eta_{\mathrm{COP}}^{\mathrm{hyb}}$ become equivalent only at weak couplings where the energetic cost $Q_{\mathrm{ec},c}$ vanishes. At strong couplings, these two COP definitions $\eta_{\mathrm{COP}}^{\mathrm{ex}}\neq \eta_{\mathrm{COP}}^{\mathrm{hyb}}$ due to a nonzero energetic cost $Q_{\mathrm{ec},c}$ for maintaining the non-Gibbsian state during the cold isothermal stroke. We thus take the presence of the second term on the right-hand-side of Eq. (\ref{eq:cop_out}) as a sign of strong coupling in the quasi-static limit. 

With the presence of an extra term on top of $\eta_{\mathrm{COP}}^{c}$, examining whether the hybrid COP definition $\eta_{\mathrm{COP}}^{\mathrm{hyb}}$ is consistent with the second law of thermodynamic becomes necessary. We note that the denominator $W_{\mathrm{ex}}$ of the extra term in Eq. (\ref{eq:cop_out}) is positive in a refrigerator, hence the sign of the extra term is fully determined by the numerator $Q_{\mathrm{ec},c}$. To analyze the sign of $Q_{\mathrm{ec},c}$, let us focus on the cold isothermal expansion stroke of the reversed Carnot cycle. During this stroke, the working substance generates work output, implying $W_{\mathrm{ex},c}<0$ and $W_c<0$. Since $|W_{\mathrm{ex},c}|$ defines the maximum work, we should have $|W_{\mathrm{ex},c}|\ge|W_c|$, or equivalently, $W_{\mathrm{ex},c}\le W_c$. We thus find $Q_{\mathrm{ec},c}=W_{\mathrm{ex},c}-W_c\le0$. Hence, the hybrid COP definition is upper bounded by the Carnot COP, rendering it a valid COP definition,
\begin{equation}\label{eq:18}
\eta_{\mathrm{COP}}^{\mathrm{hyb}}~\le~\eta_{\mathrm{COP}}^c.
\end{equation}
Here, the equality is attained only at weak couplings where $Q_{\mathrm{ec},c}=0$. At strong couplings, we generally expect $\eta_{\mathrm{COP}}^{\mathrm{hyb}}<\eta_{\mathrm{COP}}^c$.  Therefore, similar to the heat engine mode, the hybrid COP definition also unveils a negative effect of strong coupling by noting that only a weak-coupling setup can reach the Carnot COP in the quasi-static limit.

\section{Discussion and conclusion}\label{sec:5}
We remark that both the excess and hybrid definitions for the figure of merit are thermodynamically valid, and one cannot definitively exclude one definition from the two based on pure thermodynamic principles at strong couplings. Nevertheless, we recommend that the hybrid definition for the figure of merit should be used in actual applications of quantum thermal machines as it faithfully recognizes the design capability of quantum thermal machines, that is, exerting a thermodynamic useful action on the targeted external systems.

We also emphasize that the present results are specifically obtained in the quasi-static limit. Turning to the finite-time operation regime where the efficiency at maximum power is of special interest \cite{Broeck.05.PRL}, we note that non-Gibbsian states can arise even at weak couplings due to the finite-time driving fields during isothermal strokes. Hence, we cannot simply associate strong coupling effects with the emergence of the non-Gibbsian character of the system state in finite time cycles. We further remark that one can still define a nonequilibrium free energy to account for an arbitrary finite time isothermal process, however, additional dissipation induced by finite-time drivings prevents us from identifying excess heat and work from the entropy and nonequilibrium free energy changes, respectively. Extending the present framework to the finite-time regime and addressing strong coupling effects on the efficiency at maximum power therein remains an open question that warrants further investigation.

Considering that the Gibbs state is completely passive, one may naturally expect that a non-Gibbsian thermal equilibrium state has a finite ergotropy \cite{Allahverdyan.04.EPL} that allows for work extraction. To see this is indeed the case, we adopt the total ergotropy definition describing scenarios with many identical system copies, $\mathcal{E}_{\mathrm{tot}}=E_S-E_S^R$ \cite{Tirone.22.A}; Here $E_S^R=\mathrm{Tr}[H_S\rho_R]$ is the mean system energy evaluated with respect to a Gibbsian reference state $\rho_R=e^{-H_S/T_R}/Z_R$ that has the same von Neumann entropy $S$ of $\rho_S$. Since $E_S^R=F_R+T_RS$ with $F=-T_R\ln Z_R$ and $E_S=\mathcal{F}+T_RS$, we can express the total ergotropy as $\mathcal{E}_{\mathrm{tot}}=\mathcal{F}-F_R$ which vanishes only when $\rho_S$ reduces to a Gibbsian form. In the quasi-static limit, one can further replace the changes of $\mathcal{F}$ and $S$ with the excess work and heat, respectively, thereby obtaining a relation between total ergotropy change and excess thermodynamic quantities.

Furthermore, it is well-known in classical thermodynamics that a quasi-static cycle implies a vanishing total entropy production which in turn, combined with the usual first law of thermodynamics, leads to the celebrated Carnot theorem. With this result in mind, it seems that our outside view that strong coupling can suppress the maximum efficiency is incompatible with the quasi-static cycle we adopted. To clarify, we first note that delivering a definition for the total entropy production over a cycle at strong couplings faces an ambiguity as well. Following the excess definition, it is evident that the total entropy production over a cycle should read $\Sigma^{\mathrm{ex}}=-Q_{\mathrm{ex},h}/T_h-Q_{\mathrm{ex},c}/T_c$. Then the result $\eta^{\mathrm{ex}}=\eta_c$ for the heat engine follows from the combination of conditions $\eta^{\mathrm{ex}}=-W_{\mathrm{ex}}/Q_{\mathrm{ex},h}$, $W_{\mathrm{ex}}+Q_{\mathrm{ex},h}+Q_{\mathrm{ex},c}=0$ and $\Sigma^{\mathrm{ex}}=0$. However, inspired by the hybrid definition, one can argue that the total entropy production over a cycle should be defined as $\Sigma^{\mathrm{hyb}}=-Q_h/T_h-Q_c/T_c$ which quantifies the entropy change associated with the heat out of the baths. Taking the heat engine mode as an example, we have $\eta^{\mathrm{hyb}}=-W/Q_{\mathrm{ex},h}=(Q_h+Q_c)/(Q_h-Q_{\mathrm{ec},h})$ with the equality $W+Q_h+Q_c=0$ [cf. Eq. (\ref{eq:3})]. Adapting the rationale that leads to Eq. (\ref{eq:18}), we can infer $Q_h-Q_{\mathrm{ec},h}\ge Q_h$ such that $\eta^{\mathrm{hyb}}\le (Q_h+Q_c)/Q_h=\eta_c$ with the last equality obtained using the quasi-static condition $\Sigma^{\mathrm{hyb}}=0$. Hence, our result is fully compatible with the quasi-static condition.

In conclusion, we addressed strong-coupling quantum Carnot thermal machines in the quasi-static limit with a pure analytical treatment, with the consideration that it is challenging to numerically implement a proper quasi-static thermodynamic cycle. We revealed that the quantum thermal machine allows for multiple definitions for the figure of merit at strong couplings. As a result, one cannot reach conclusive results regarding the strong coupling effect on maximum efficiency. Specifically, we proposed two possible definitions, dubbed excess and hybrid ones. We have analytically shown that the excess definition which just considers the energetics of the working substance exactly recovers the Carnot limit at strong couplings, thereby implying that there is no strong coupling effect on the maximum efficiency. In contrast, we demonstrated that the hybrid definition which evaluates the gain from the perspective of the external system upon which the quantum thermal machine acts predicts a suppression effect of the strong coupling, rendering the Carnot limit unattainable at strong couplings. Notably, the latter result extends that in Ref. \cite{Liu.22.PRE} to regimes beyond the linear response. Our results thus point out that it is necessary to first gauge the definition for the figure of merit before trying to elucidate the strong coupling effect on the performance. We believe this work highlights an important issue in the field and will lead to further investigations on the nature of strong-coupling quantum thermodynamics and quantum thermal machines.

\begin{acknowledgments}
J.L. acknowledges support from the National Natural Science
Foundation of China (Grant No. 12205179), the Shanghai Pujiang Program
(Grant No. 22PJ1403900) and start-up funding of Shanghai University.
\end{acknowledgments}


\begin{thebibliography}{57}%
\makeatletter
\providecommand \@ifxundefined [1]{%
 \@ifx{#1\undefined}
}%
\providecommand \@ifnum [1]{%
 \ifnum #1\expandafter \@firstoftwo
 \else \expandafter \@secondoftwo
 \fi
}%
\providecommand \@ifx [1]{%
 \ifx #1\expandafter \@firstoftwo
 \else \expandafter \@secondoftwo
 \fi
}%
\providecommand \natexlab [1]{#1}%
\providecommand \enquote  [1]{``#1''}%
\providecommand \bibnamefont  [1]{#1}%
\providecommand \bibfnamefont [1]{#1}%
\providecommand \citenamefont [1]{#1}%
\providecommand \href@noop [0]{\@secondoftwo}%
\providecommand \href [0]{\begingroup \@sanitize@url \@href}%
\providecommand \@href[1]{\@@startlink{#1}\@@href}%
\providecommand \@@href[1]{\endgroup#1\@@endlink}%
\providecommand \@sanitize@url [0]{\catcode `\\12\catcode `\$12\catcode
  `\&12\catcode `\#12\catcode `\^12\catcode `\_12\catcode `\%12\relax}%
\providecommand \@@startlink[1]{}%
\providecommand \@@endlink[0]{}%
\providecommand \url  [0]{\begingroup\@sanitize@url \@url }%
\providecommand \@url [1]{\endgroup\@href {#1}{\urlprefix }}%
\providecommand \urlprefix  [0]{URL }%
\providecommand \Eprint [0]{\href }%
\providecommand \doibase [0]{http://dx.doi.org/}%
\providecommand \selectlanguage [0]{\@gobble}%
\providecommand \bibinfo  [0]{\@secondoftwo}%
\providecommand \bibfield  [0]{\@secondoftwo}%
\providecommand \translation [1]{[#1]}%
\providecommand \BibitemOpen [0]{}%
\providecommand \bibitemStop [0]{}%
\providecommand \bibitemNoStop [0]{.\EOS\space}%
\providecommand \EOS [0]{\spacefactor3000\relax}%
\providecommand \BibitemShut  [1]{\csname bibitem#1\endcsname}%
\let\auto@bib@innerbib\@empty
\bibitem [{\citenamefont {Benenti}\ \emph {et~al.}(2017)\citenamefont
  {Benenti}, \citenamefont {Casati}, \citenamefont {Saito},\ and\ \citenamefont
  {Whitney}}]{Benenti.17.PR}%
  \BibitemOpen
  \bibfield  {author} {\bibinfo {author} {\bibfnamefont {G.}~\bibnamefont
  {Benenti}}, \bibinfo {author} {\bibfnamefont {G.}~\bibnamefont {Casati}},
  \bibinfo {author} {\bibfnamefont {K.}~\bibnamefont {Saito}}, \ and\ \bibinfo
  {author} {\bibfnamefont {R.}~\bibnamefont {Whitney}},\ }\bibfield  {title}
  {\enquote {\bibinfo {title} {Fundamental aspects of steady-state conversion
  of heat to work at the nanoscale},}\ }\href {\doibase
  https://doi.org/10.1016/j.physrep.2017.05.008} {\bibfield  {journal}
  {\bibinfo  {journal} {Phys. Rep.}\ }\textbf {\bibinfo {volume} {694}},\
  \bibinfo {pages} {1} (\bibinfo {year} {2017})}\BibitemShut {NoStop}%
\bibitem [{\citenamefont {Mitchison}(2019)}]{Mitchison.19.CP}%
  \BibitemOpen
  \bibfield  {author} {\bibinfo {author} {\bibfnamefont {M.~T.}\ \bibnamefont
  {Mitchison}},\ }\bibfield  {title} {\enquote {\bibinfo {title} {Quantum
  thermal absorption machines: refrigerators, engines and clocks},}\ }\href
  {\doibase 10.1080/00107514.2019.1631555} {\bibfield  {journal} {\bibinfo
  {journal} {Contemporary Phys.}\ }\textbf {\bibinfo {volume} {60}},\ \bibinfo
  {pages} {164} (\bibinfo {year} {2019})}\BibitemShut {NoStop}%
\bibitem [{\citenamefont {Myers}\ \emph {et~al.}(2022)\citenamefont {Myers},
  \citenamefont {Abah},\ and\ \citenamefont {Deffner}}]{Myers.22.AVSQS}%
  \BibitemOpen
  \bibfield  {author} {\bibinfo {author} {\bibfnamefont {N.}~\bibnamefont
  {Myers}}, \bibinfo {author} {\bibfnamefont {O.}~\bibnamefont {Abah}}, \ and\
  \bibinfo {author} {\bibfnamefont {S.}~\bibnamefont {Deffner}},\ }\bibfield
  {title} {\enquote {\bibinfo {title} {{Quantum thermodynamic devices: From
  theoretical proposals to experimental reality}},}\ }\href {\doibase
  10.1116/5.0083192} {\bibfield  {journal} {\bibinfo  {journal} {AVS Quantum
  Sci.}\ }\textbf {\bibinfo {volume} {4}},\ \bibinfo {pages} {027101} (\bibinfo
  {year} {2022})}\BibitemShut {NoStop}%
\bibitem [{\citenamefont {Buffoni}\ \emph {et~al.}(2019)\citenamefont
  {Buffoni}, \citenamefont {Solfanelli}, \citenamefont {Verrucchi},
  \citenamefont {Cuccoli},\ and\ \citenamefont {Campisi}}]{Buffoni.19.PRL}%
  \BibitemOpen
  \bibfield  {author} {\bibinfo {author} {\bibfnamefont {L.}~\bibnamefont
  {Buffoni}}, \bibinfo {author} {\bibfnamefont {A.}~\bibnamefont {Solfanelli}},
  \bibinfo {author} {\bibfnamefont {P.}~\bibnamefont {Verrucchi}}, \bibinfo
  {author} {\bibfnamefont {A.}~\bibnamefont {Cuccoli}}, \ and\ \bibinfo
  {author} {\bibfnamefont {M.}~\bibnamefont {Campisi}},\ }\bibfield  {title}
  {\enquote {\bibinfo {title} {Quantum measurement cooling},}\ }\href {\doibase
  10.1103/PhysRevLett.122.070603} {\bibfield  {journal} {\bibinfo  {journal}
  {Phys. Rev. Lett.}\ }\textbf {\bibinfo {volume} {122}},\ \bibinfo {pages}
  {070603} (\bibinfo {year} {2019})}\BibitemShut {NoStop}%
\bibitem [{\citenamefont {Carrega}\ \emph {et~al.}(2022)\citenamefont
  {Carrega}, \citenamefont {Cangemi}, \citenamefont {De~Filippis},
  \citenamefont {Cataudella}, \citenamefont {Benenti},\ and\ \citenamefont
  {Sassetti}}]{Carrega.22.PRXQ}%
  \BibitemOpen
  \bibfield  {author} {\bibinfo {author} {\bibfnamefont {M.}~\bibnamefont
  {Carrega}}, \bibinfo {author} {\bibfnamefont {L.}~\bibnamefont {Cangemi}},
  \bibinfo {author} {\bibfnamefont {G.}~\bibnamefont {De~Filippis}}, \bibinfo
  {author} {\bibfnamefont {V.}~\bibnamefont {Cataudella}}, \bibinfo {author}
  {\bibfnamefont {G.}~\bibnamefont {Benenti}}, \ and\ \bibinfo {author}
  {\bibfnamefont {M.}~\bibnamefont {Sassetti}},\ }\bibfield  {title} {\enquote
  {\bibinfo {title} {Engineering dynamical couplings for quantum thermodynamic
  tasks},}\ }\href {\doibase 10.1103/PRXQuantum.3.010323} {\bibfield  {journal}
  {\bibinfo  {journal} {PRX Quantum}\ }\textbf {\bibinfo {volume} {3}},\
  \bibinfo {pages} {010323} (\bibinfo {year} {2022})}\BibitemShut {NoStop}%
\bibitem [{\citenamefont {Ji}\ \emph {et~al.}(2022)\citenamefont {Ji},
  \citenamefont {Chai}, \citenamefont {Wang}, \citenamefont {Guo},
  \citenamefont {Rong}, \citenamefont {Shi}, \citenamefont {Ren}, \citenamefont
  {Wang},\ and\ \citenamefont {Du}}]{JiW.22.PRL}%
  \BibitemOpen
  \bibfield  {author} {\bibinfo {author} {\bibfnamefont {W.}~\bibnamefont
  {Ji}}, \bibinfo {author} {\bibfnamefont {Z.}~\bibnamefont {Chai}}, \bibinfo
  {author} {\bibfnamefont {M.}~\bibnamefont {Wang}}, \bibinfo {author}
  {\bibfnamefont {Y.}~\bibnamefont {Guo}}, \bibinfo {author} {\bibfnamefont
  {X.}~\bibnamefont {Rong}}, \bibinfo {author} {\bibfnamefont {F.}~\bibnamefont
  {Shi}}, \bibinfo {author} {\bibfnamefont {C.}~\bibnamefont {Ren}}, \bibinfo
  {author} {\bibfnamefont {Y.}~\bibnamefont {Wang}}, \ and\ \bibinfo {author}
  {\bibfnamefont {J.}~\bibnamefont {Du}},\ }\bibfield  {title} {\enquote
  {\bibinfo {title} {Spin quantum heat engine quantified by quantum
  steering},}\ }\href {\doibase 10.1103/PhysRevLett.128.090602} {\bibfield
  {journal} {\bibinfo  {journal} {Phys. Rev. Lett.}\ }\textbf {\bibinfo
  {volume} {128}},\ \bibinfo {pages} {090602} (\bibinfo {year}
  {2022})}\BibitemShut {NoStop}%
\bibitem [{\citenamefont {Lu}\ \emph {et~al.}(2023)\citenamefont {Lu},
  \citenamefont {Wang}, \citenamefont {Wang}, \citenamefont {Peng},
  \citenamefont {Wang},\ and\ \citenamefont {Jiang}}]{Lu.23.PRB}%
  \BibitemOpen
  \bibfield  {author} {\bibinfo {author} {\bibfnamefont {J.}~\bibnamefont
  {Lu}}, \bibinfo {author} {\bibfnamefont {Z.}~\bibnamefont {Wang}}, \bibinfo
  {author} {\bibfnamefont {R.}~\bibnamefont {Wang}}, \bibinfo {author}
  {\bibfnamefont {J.}~\bibnamefont {Peng}}, \bibinfo {author} {\bibfnamefont
  {C.}~\bibnamefont {Wang}}, \ and\ \bibinfo {author} {\bibfnamefont
  {J.}~\bibnamefont {Jiang}},\ }\bibfield  {title} {\enquote {\bibinfo {title}
  {Multitask quantum thermal machines and cooperative effects},}\ }\href
  {\doibase 10.1103/PhysRevB.107.075428} {\bibfield  {journal} {\bibinfo
  {journal} {Phys. Rev. B}\ }\textbf {\bibinfo {volume} {107}},\ \bibinfo
  {pages} {075428} (\bibinfo {year} {2023})}\BibitemShut {NoStop}%
\bibitem [{\citenamefont {Quan}\ \emph {et~al.}(2007)\citenamefont {Quan},
  \citenamefont {Liu}, \citenamefont {Sun},\ and\ \citenamefont
  {Nori}}]{Quan.07.PRE}%
  \BibitemOpen
  \bibfield  {author} {\bibinfo {author} {\bibfnamefont {H.~T.}\ \bibnamefont
  {Quan}}, \bibinfo {author} {\bibfnamefont {Yu-xi}\ \bibnamefont {Liu}},
  \bibinfo {author} {\bibfnamefont {C.~P.}\ \bibnamefont {Sun}}, \ and\
  \bibinfo {author} {\bibfnamefont {F.}~\bibnamefont {Nori}},\ }\bibfield
  {title} {\enquote {\bibinfo {title} {Quantum thermodynamic cycles and quantum
  heat engines},}\ }\href {\doibase 10.1103/PhysRevE.76.031105} {\bibfield
  {journal} {\bibinfo  {journal} {Phys. Rev. E}\ }\textbf {\bibinfo {volume}
  {76}},\ \bibinfo {pages} {031105} (\bibinfo {year} {2007})}\BibitemShut
  {NoStop}%
\bibitem [{\citenamefont {Brandner}\ \emph {et~al.}(2015)\citenamefont
  {Brandner}, \citenamefont {Saito},\ and\ \citenamefont
  {Seifert}}]{Brandner.15.PRX}%
  \BibitemOpen
  \bibfield  {author} {\bibinfo {author} {\bibfnamefont {K.}~\bibnamefont
  {Brandner}}, \bibinfo {author} {\bibfnamefont {K.}~\bibnamefont {Saito}}, \
  and\ \bibinfo {author} {\bibfnamefont {U.}~\bibnamefont {Seifert}},\
  }\bibfield  {title} {\enquote {\bibinfo {title} {Thermodynamics of micro- and
  nano-systems driven by periodic temperature variations},}\ }\href {\doibase
  10.1103/PhysRevX.5.031019} {\bibfield  {journal} {\bibinfo  {journal} {Phys.
  Rev. X}\ }\textbf {\bibinfo {volume} {5}},\ \bibinfo {pages} {031019}
  (\bibinfo {year} {2015})}\BibitemShut {NoStop}%
\bibitem [{\citenamefont {Proesmans}\ \emph {et~al.}(2016)\citenamefont
  {Proesmans}, \citenamefont {Dreher}, \citenamefont {Gavrilov}, \citenamefont
  {Bechhoefer},\ and\ \citenamefont {Van~den Broeck}}]{Proesmans.16.PRX}%
  \BibitemOpen
  \bibfield  {author} {\bibinfo {author} {\bibfnamefont {K.}~\bibnamefont
  {Proesmans}}, \bibinfo {author} {\bibfnamefont {Y.}~\bibnamefont {Dreher}},
  \bibinfo {author} {\bibfnamefont {M.}~\bibnamefont {Gavrilov}}, \bibinfo
  {author} {\bibfnamefont {J.}~\bibnamefont {Bechhoefer}}, \ and\ \bibinfo
  {author} {\bibfnamefont {C.}~\bibnamefont {Van~den Broeck}},\ }\bibfield
  {title} {\enquote {\bibinfo {title} {Brownian duet: A novel tale of
  thermodynamic efficiency},}\ }\href {\doibase 10.1103/PhysRevX.6.041010}
  {\bibfield  {journal} {\bibinfo  {journal} {Phys. Rev. X}\ }\textbf {\bibinfo
  {volume} {6}},\ \bibinfo {pages} {041010} (\bibinfo {year}
  {2016})}\BibitemShut {NoStop}%
\bibitem [{\citenamefont {Newman}\ \emph {et~al.}(2017)\citenamefont {Newman},
  \citenamefont {Mintert},\ and\ \citenamefont {Nazir}}]{Newman.17.PRE}%
  \BibitemOpen
  \bibfield  {author} {\bibinfo {author} {\bibfnamefont {D.}~\bibnamefont
  {Newman}}, \bibinfo {author} {\bibfnamefont {F.}~\bibnamefont {Mintert}}, \
  and\ \bibinfo {author} {\bibfnamefont {A.}~\bibnamefont {Nazir}},\ }\bibfield
   {title} {\enquote {\bibinfo {title} {Performance of a quantum heat engine at
  strong reservoir coupling},}\ }\href {\doibase 10.1103/PhysRevE.95.032139}
  {\bibfield  {journal} {\bibinfo  {journal} {Phys. Rev. E}\ }\textbf {\bibinfo
  {volume} {95}},\ \bibinfo {pages} {032139} (\bibinfo {year}
  {2017})}\BibitemShut {NoStop}%
\bibitem [{\citenamefont {Watanabe}\ \emph {et~al.}(2017)\citenamefont
  {Watanabe}, \citenamefont {Venkatesh}, \citenamefont {Talkner},\ and\
  \citenamefont {del Campo}}]{Watanabe.17.PRL}%
  \BibitemOpen
  \bibfield  {author} {\bibinfo {author} {\bibfnamefont {G.}~\bibnamefont
  {Watanabe}}, \bibinfo {author} {\bibfnamefont {B.}~\bibnamefont {Venkatesh}},
  \bibinfo {author} {\bibfnamefont {P.}~\bibnamefont {Talkner}}, \ and\
  \bibinfo {author} {\bibfnamefont {A.}~\bibnamefont {del Campo}},\ }\bibfield
  {title} {\enquote {\bibinfo {title} {Quantum performance of thermal machines
  over many cycles},}\ }\href {\doibase 10.1103/PhysRevLett.118.050601}
  {\bibfield  {journal} {\bibinfo  {journal} {Phys. Rev. Lett.}\ }\textbf
  {\bibinfo {volume} {118}},\ \bibinfo {pages} {050601} (\bibinfo {year}
  {2017})}\BibitemShut {NoStop}%
\bibitem [{\citenamefont {Perarnau-Llobet}\ \emph {et~al.}(2018)\citenamefont
  {Perarnau-Llobet}, \citenamefont {Wilming}, \citenamefont {Riera},
  \citenamefont {Gallego},\ and\ \citenamefont
  {Eisert}}]{Perarnau-Llobet.18.PRL}%
  \BibitemOpen
  \bibfield  {author} {\bibinfo {author} {\bibfnamefont {M.}~\bibnamefont
  {Perarnau-Llobet}}, \bibinfo {author} {\bibfnamefont {H.}~\bibnamefont
  {Wilming}}, \bibinfo {author} {\bibfnamefont {A.}~\bibnamefont {Riera}},
  \bibinfo {author} {\bibfnamefont {R.}~\bibnamefont {Gallego}}, \ and\
  \bibinfo {author} {\bibfnamefont {J.}~\bibnamefont {Eisert}},\ }\bibfield
  {title} {\enquote {\bibinfo {title} {Strong coupling corrections in quantum
  thermodynamics},}\ }\href {\doibase 10.1103/PhysRevLett.120.120602}
  {\bibfield  {journal} {\bibinfo  {journal} {Phys. Rev. Lett.}\ }\textbf
  {\bibinfo {volume} {120}},\ \bibinfo {pages} {120602} (\bibinfo {year}
  {2018})}\BibitemShut {NoStop}%
\bibitem [{\citenamefont {Mohammady}\ and\ \citenamefont
  {Romito}(2019)}]{Mohammady.19.PRE}%
  \BibitemOpen
  \bibfield  {author} {\bibinfo {author} {\bibfnamefont {M.}~\bibnamefont
  {Mohammady}}\ and\ \bibinfo {author} {\bibfnamefont {A.}~\bibnamefont
  {Romito}},\ }\bibfield  {title} {\enquote {\bibinfo {title} {Efficiency of a
  cyclic quantum heat engine with finite-size baths},}\ }\href {\doibase
  10.1103/PhysRevE.100.012122} {\bibfield  {journal} {\bibinfo  {journal}
  {Phys. Rev. E}\ }\textbf {\bibinfo {volume} {100}},\ \bibinfo {pages}
  {012122} (\bibinfo {year} {2019})}\BibitemShut {NoStop}%
\bibitem [{\citenamefont {Ono}\ \emph {et~al.}(2020)\citenamefont {Ono},
  \citenamefont {Shevchenko}, \citenamefont {Mori}, \citenamefont {Moriyama},\
  and\ \citenamefont {Nori}}]{Ono.20.PRL}%
  \BibitemOpen
  \bibfield  {author} {\bibinfo {author} {\bibfnamefont {K.}~\bibnamefont
  {Ono}}, \bibinfo {author} {\bibfnamefont {S.~N.}\ \bibnamefont {Shevchenko}},
  \bibinfo {author} {\bibfnamefont {T.}~\bibnamefont {Mori}}, \bibinfo {author}
  {\bibfnamefont {S.}~\bibnamefont {Moriyama}}, \ and\ \bibinfo {author}
  {\bibfnamefont {Franco}\ \bibnamefont {Nori}},\ }\bibfield  {title} {\enquote
  {\bibinfo {title} {Analog of a quantum heat engine using a single-spin
  qubit},}\ }\href {\doibase 10.1103/PhysRevLett.125.166802} {\bibfield
  {journal} {\bibinfo  {journal} {Phys. Rev. Lett.}\ }\textbf {\bibinfo
  {volume} {125}},\ \bibinfo {pages} {166802} (\bibinfo {year}
  {2020})}\BibitemShut {NoStop}%
\bibitem [{\citenamefont {Strasberg}\ \emph
  {et~al.}(2021{\natexlab{a}})\citenamefont {Strasberg}, \citenamefont
  {W\"achtler},\ and\ \citenamefont {Schaller}}]{Strasberg.21.PRL}%
  \BibitemOpen
  \bibfield  {author} {\bibinfo {author} {\bibfnamefont {P.}~\bibnamefont
  {Strasberg}}, \bibinfo {author} {\bibfnamefont {C.}~\bibnamefont
  {W\"achtler}}, \ and\ \bibinfo {author} {\bibfnamefont {G.}~\bibnamefont
  {Schaller}},\ }\bibfield  {title} {\enquote {\bibinfo {title} {Autonomous
  implementation of thermodynamic cycles at the nanoscale},}\ }\href {\doibase
  10.1103/PhysRevLett.126.180605} {\bibfield  {journal} {\bibinfo  {journal}
  {Phys. Rev. Lett.}\ }\textbf {\bibinfo {volume} {126}},\ \bibinfo {pages}
  {180605} (\bibinfo {year} {2021}{\natexlab{a}})}\BibitemShut {NoStop}%
\bibitem [{\citenamefont {Strasberg}\ \emph
  {et~al.}(2021{\natexlab{b}})\citenamefont {Strasberg}, \citenamefont
  {D\'{\i}az},\ and\ \citenamefont {Riera-Campeny}}]{Strasberg.21.PRE}%
  \BibitemOpen
  \bibfield  {author} {\bibinfo {author} {\bibfnamefont {P.}~\bibnamefont
  {Strasberg}}, \bibinfo {author} {\bibfnamefont {M.}~\bibnamefont
  {D\'{\i}az}}, \ and\ \bibinfo {author} {\bibfnamefont {A.}~\bibnamefont
  {Riera-Campeny}},\ }\bibfield  {title} {\enquote {\bibinfo {title} {Clausius
  inequality for finite baths reveals universal efficiency improvements},}\
  }\href {\doibase 10.1103/PhysRevE.104.L022103} {\bibfield  {journal}
  {\bibinfo  {journal} {Phys. Rev. E}\ }\textbf {\bibinfo {volume} {104}},\
  \bibinfo {pages} {L022103} (\bibinfo {year}
  {2021}{\natexlab{b}})}\BibitemShut {NoStop}%
\bibitem [{\citenamefont {Liu}\ \emph {et~al.}(2021)\citenamefont {Liu},
  \citenamefont {Jung},\ and\ \citenamefont {Segal}}]{Liu.21.PRL}%
  \BibitemOpen
  \bibfield  {author} {\bibinfo {author} {\bibfnamefont {J.}~\bibnamefont
  {Liu}}, \bibinfo {author} {\bibfnamefont {K.}~\bibnamefont {Jung}}, \ and\
  \bibinfo {author} {\bibfnamefont {D.}~\bibnamefont {Segal}},\ }\bibfield
  {title} {\enquote {\bibinfo {title} {Periodically driven quantum thermal
  machines from warming up to limit cycle},}\ }\href {\doibase
  10.1103/PhysRevLett.127.200602} {\bibfield  {journal} {\bibinfo  {journal}
  {Phys. Rev. Lett.}\ }\textbf {\bibinfo {volume} {127}},\ \bibinfo {pages}
  {200602} (\bibinfo {year} {2021})}\BibitemShut {NoStop}%
\bibitem [{\citenamefont {Abah}\ \emph {et~al.}(2012)\citenamefont {Abah},
  \citenamefont {Ro\ss{}nagel}, \citenamefont {Jacob}, \citenamefont {Deffner},
  \citenamefont {Schmidt-Kaler}, \citenamefont {Singer},\ and\ \citenamefont
  {Lutz}}]{Abah.12.PRL}%
  \BibitemOpen
  \bibfield  {author} {\bibinfo {author} {\bibfnamefont {O.}~\bibnamefont
  {Abah}}, \bibinfo {author} {\bibfnamefont {J.}~\bibnamefont {Ro\ss{}nagel}},
  \bibinfo {author} {\bibfnamefont {G.}~\bibnamefont {Jacob}}, \bibinfo
  {author} {\bibfnamefont {S.}~\bibnamefont {Deffner}}, \bibinfo {author}
  {\bibfnamefont {F.}~\bibnamefont {Schmidt-Kaler}}, \bibinfo {author}
  {\bibfnamefont {K.}~\bibnamefont {Singer}}, \ and\ \bibinfo {author}
  {\bibfnamefont {E.}~\bibnamefont {Lutz}},\ }\bibfield  {title} {\enquote
  {\bibinfo {title} {Single-ion heat engine at maximum power},}\ }\href
  {\doibase 10.1103/PhysRevLett.109.203006} {\bibfield  {journal} {\bibinfo
  {journal} {Phys. Rev. Lett.}\ }\textbf {\bibinfo {volume} {109}},\ \bibinfo
  {pages} {203006} (\bibinfo {year} {2012})}\BibitemShut {NoStop}%
\bibitem [{\citenamefont {Ro\ss{}nagel}\ \emph {et~al.}(2014)\citenamefont
  {Ro\ss{}nagel}, \citenamefont {Abah}, \citenamefont {Schmidt-Kaler},
  \citenamefont {Singer},\ and\ \citenamefont {Lutz}}]{Rossnagel.14.PRL}%
  \BibitemOpen
  \bibfield  {author} {\bibinfo {author} {\bibfnamefont {J.}~\bibnamefont
  {Ro\ss{}nagel}}, \bibinfo {author} {\bibfnamefont {O.}~\bibnamefont {Abah}},
  \bibinfo {author} {\bibfnamefont {F.}~\bibnamefont {Schmidt-Kaler}}, \bibinfo
  {author} {\bibfnamefont {K.}~\bibnamefont {Singer}}, \ and\ \bibinfo {author}
  {\bibfnamefont {E.}~\bibnamefont {Lutz}},\ }\bibfield  {title} {\enquote
  {\bibinfo {title} {Nanoscale heat engine beyond the carnot limit},}\ }\href
  {\doibase 10.1103/PhysRevLett.112.030602} {\bibfield  {journal} {\bibinfo
  {journal} {Phys. Rev. Lett.}\ }\textbf {\bibinfo {volume} {112}},\ \bibinfo
  {pages} {030602} (\bibinfo {year} {2014})}\BibitemShut {NoStop}%
\bibitem [{\citenamefont {Ro\ss{}nagel}\ \emph {et~al.}(2016)\citenamefont
  {Ro\ss{}nagel}, \citenamefont {Dawkins}, \citenamefont {Tolazzi},
  \citenamefont {Abah}, \citenamefont {Lutz}, \citenamefont {Schmidt-Kaler},\
  and\ \citenamefont {Singer}}]{Robnagel.16.S}%
  \BibitemOpen
  \bibfield  {author} {\bibinfo {author} {\bibfnamefont {J.}~\bibnamefont
  {Ro\ss{}nagel}}, \bibinfo {author} {\bibfnamefont {Samuel~T.}\ \bibnamefont
  {Dawkins}}, \bibinfo {author} {\bibfnamefont {K.~N.}\ \bibnamefont
  {Tolazzi}}, \bibinfo {author} {\bibfnamefont {O.}~\bibnamefont {Abah}},
  \bibinfo {author} {\bibfnamefont {E.}~\bibnamefont {Lutz}}, \bibinfo {author}
  {\bibfnamefont {F.}~\bibnamefont {Schmidt-Kaler}}, \ and\ \bibinfo {author}
  {\bibfnamefont {K.}~\bibnamefont {Singer}},\ }\bibfield  {title} {\enquote
  {\bibinfo {title} {A single-atom heat engine},}\ }\href
  {http://science.sciencemag.org/content/352/6283/325.abstract} {\bibfield
  {journal} {\bibinfo  {journal} {Science}\ }\textbf {\bibinfo {volume}
  {352}},\ \bibinfo {pages} {325} (\bibinfo {year} {2016})}\BibitemShut
  {NoStop}%
\bibitem [{\citenamefont {Krishnamurthy}\ \emph {et~al.}(2016)\citenamefont
  {Krishnamurthy}, \citenamefont {Ghosh}, \citenamefont {Chatterji},
  \citenamefont {Ganapathy},\ and\ \citenamefont {Sood}}]{Sood.16.NP}%
  \BibitemOpen
  \bibfield  {author} {\bibinfo {author} {\bibfnamefont {S.}~\bibnamefont
  {Krishnamurthy}}, \bibinfo {author} {\bibfnamefont {S.}~\bibnamefont
  {Ghosh}}, \bibinfo {author} {\bibfnamefont {D.}~\bibnamefont {Chatterji}},
  \bibinfo {author} {\bibfnamefont {R.}~\bibnamefont {Ganapathy}}, \ and\
  \bibinfo {author} {\bibfnamefont {A.~K.}\ \bibnamefont {Sood}},\ }\bibfield
  {title} {\enquote {\bibinfo {title} {A micrometre-sized heat engine operating
  between bacterial reservoirs},}\ }\href {https://doi.org/10.1038/nphys3870}
  {\bibfield  {journal} {\bibinfo  {journal} {Nat. Phys.}\ }\textbf {\bibinfo
  {volume} {12}},\ \bibinfo {pages} {1134} (\bibinfo {year}
  {2016})}\BibitemShut {NoStop}%
\bibitem [{\citenamefont {Peterson}\ \emph {et~al.}(2019)\citenamefont
  {Peterson}, \citenamefont {Batalh\~ao}, \citenamefont {Herrera},
  \citenamefont {Souza}, \citenamefont {Sarthour}, \citenamefont {Oliveira},\
  and\ \citenamefont {Serra}}]{Peterson.19.PRL}%
  \BibitemOpen
  \bibfield  {author} {\bibinfo {author} {\bibfnamefont {J.~P.~S.}\
  \bibnamefont {Peterson}}, \bibinfo {author} {\bibfnamefont {T.~B.}\
  \bibnamefont {Batalh\~ao}}, \bibinfo {author} {\bibfnamefont
  {M.}~\bibnamefont {Herrera}}, \bibinfo {author} {\bibfnamefont {A.~M.}\
  \bibnamefont {Souza}}, \bibinfo {author} {\bibfnamefont {R.~S.}\ \bibnamefont
  {Sarthour}}, \bibinfo {author} {\bibfnamefont {I.~S.}\ \bibnamefont
  {Oliveira}}, \ and\ \bibinfo {author} {\bibfnamefont {R.~M.}\ \bibnamefont
  {Serra}},\ }\bibfield  {title} {\enquote {\bibinfo {title} {Experimental
  characterization of a spin quantum heat engine},}\ }\href {\doibase
  10.1103/PhysRevLett.123.240601} {\bibfield  {journal} {\bibinfo  {journal}
  {Phys. Rev. Lett.}\ }\textbf {\bibinfo {volume} {123}},\ \bibinfo {pages}
  {240601} (\bibinfo {year} {2019})}\BibitemShut {NoStop}%
\bibitem [{\citenamefont {Klatzow}\ \emph {et~al.}(2019)\citenamefont
  {Klatzow}, \citenamefont {Becker}, \citenamefont {Ledingham}, \citenamefont
  {Weinzetl}, \citenamefont {Kaczmarek}, \citenamefont {Saunders},
  \citenamefont {Nunn}, \citenamefont {Walmsley}, \citenamefont {Uzdin},\ and\
  \citenamefont {Poem}}]{Klatzow.19.PRL}%
  \BibitemOpen
  \bibfield  {author} {\bibinfo {author} {\bibfnamefont {J.}~\bibnamefont
  {Klatzow}}, \bibinfo {author} {\bibfnamefont {J.~N.}\ \bibnamefont {Becker}},
  \bibinfo {author} {\bibfnamefont {P.~M.}\ \bibnamefont {Ledingham}}, \bibinfo
  {author} {\bibfnamefont {C.}~\bibnamefont {Weinzetl}}, \bibinfo {author}
  {\bibfnamefont {K.~T.}\ \bibnamefont {Kaczmarek}}, \bibinfo {author}
  {\bibfnamefont {D.~J.}\ \bibnamefont {Saunders}}, \bibinfo {author}
  {\bibfnamefont {J.}~\bibnamefont {Nunn}}, \bibinfo {author} {\bibfnamefont
  {I.~A.}\ \bibnamefont {Walmsley}}, \bibinfo {author} {\bibfnamefont
  {R.}~\bibnamefont {Uzdin}}, \ and\ \bibinfo {author} {\bibfnamefont
  {E.}~\bibnamefont {Poem}},\ }\bibfield  {title} {\enquote {\bibinfo {title}
  {Experimental demonstration of quantum effects in the operation of
  microscopic heat engines},}\ }\href {\doibase 10.1103/PhysRevLett.122.110601}
  {\bibfield  {journal} {\bibinfo  {journal} {Phys. Rev. Lett.}\ }\textbf
  {\bibinfo {volume} {122}},\ \bibinfo {pages} {110601} (\bibinfo {year}
  {2019})}\BibitemShut {NoStop}%
\bibitem [{\citenamefont {de~Assis}\ \emph {et~al.}(2019)\citenamefont
  {de~Assis}, \citenamefont {de~Mendon\ifmmode~\mbox{\c{c}}\else \c{c}\fi{}a},
  \citenamefont {Villas-Boas}, \citenamefont {de~Souza}, \citenamefont
  {Sarthour}, \citenamefont {Oliveira},\ and\ \citenamefont
  {de~Almeida}}]{Assis.19.PRL}%
  \BibitemOpen
  \bibfield  {author} {\bibinfo {author} {\bibfnamefont {R.~J.}\ \bibnamefont
  {de~Assis}}, \bibinfo {author} {\bibfnamefont {T.}~\bibnamefont
  {de~Mendon\ifmmode~\mbox{\c{c}}\else \c{c}\fi{}a}}, \bibinfo {author}
  {\bibfnamefont {C.}~\bibnamefont {Villas-Boas}}, \bibinfo {author}
  {\bibfnamefont {A.}~\bibnamefont {de~Souza}}, \bibinfo {author}
  {\bibfnamefont {R.}~\bibnamefont {Sarthour}}, \bibinfo {author}
  {\bibfnamefont {I.}~\bibnamefont {Oliveira}}, \ and\ \bibinfo {author}
  {\bibfnamefont {N.}~\bibnamefont {de~Almeida}},\ }\bibfield  {title}
  {\enquote {\bibinfo {title} {Efficiency of a quantum otto heat engine
  operating under a reservoir at effective negative temperatures},}\ }\href
  {\doibase 10.1103/PhysRevLett.122.240602} {\bibfield  {journal} {\bibinfo
  {journal} {Phys. Rev. Lett.}\ }\textbf {\bibinfo {volume} {122}},\ \bibinfo
  {pages} {240602} (\bibinfo {year} {2019})}\BibitemShut {NoStop}%
\bibitem [{\citenamefont {Pekola}\ and\ \citenamefont
  {Khaymovich}(2019)}]{Pekola.19.ARCMP}%
  \BibitemOpen
  \bibfield  {author} {\bibinfo {author} {\bibfnamefont {J.}~\bibnamefont
  {Pekola}}\ and\ \bibinfo {author} {\bibfnamefont {I.}~\bibnamefont
  {Khaymovich}},\ }\bibfield  {title} {\enquote {\bibinfo {title}
  {Thermodynamics in single-electron circuits and superconducting qubits},}\
  }\href {\doibase 10.1146/annurev-conmatphys-033117-054120} {\bibfield
  {journal} {\bibinfo  {journal} {Annu. Rev. Condens. Matter Phys.}\ }\textbf
  {\bibinfo {volume} {10}},\ \bibinfo {pages} {193} (\bibinfo {year}
  {2019})}\BibitemShut {NoStop}%
\bibitem [{\citenamefont {von Lindenfels}\ \emph {et~al.}(2019)\citenamefont
  {von Lindenfels}, \citenamefont {Gr\"ab}, \citenamefont {Schmiegelow},
  \citenamefont {Kaushal}, \citenamefont {Schulz}, \citenamefont {Mitchison},
  \citenamefont {Goold}, \citenamefont {Schmidt-Kaler},\ and\ \citenamefont
  {Poschinger}}]{Lindenfels.19.PRL}%
  \BibitemOpen
  \bibfield  {author} {\bibinfo {author} {\bibfnamefont {D.}~\bibnamefont {von
  Lindenfels}}, \bibinfo {author} {\bibfnamefont {O.}~\bibnamefont {Gr\"ab}},
  \bibinfo {author} {\bibfnamefont {C.}~\bibnamefont {Schmiegelow}}, \bibinfo
  {author} {\bibfnamefont {V.}~\bibnamefont {Kaushal}}, \bibinfo {author}
  {\bibfnamefont {J.}~\bibnamefont {Schulz}}, \bibinfo {author} {\bibfnamefont
  {M.}~\bibnamefont {Mitchison}}, \bibinfo {author} {\bibfnamefont
  {J.}~\bibnamefont {Goold}}, \bibinfo {author} {\bibfnamefont
  {F.}~\bibnamefont {Schmidt-Kaler}}, \ and\ \bibinfo {author} {\bibfnamefont
  {U.}~\bibnamefont {Poschinger}},\ }\bibfield  {title} {\enquote {\bibinfo
  {title} {Spin heat engine coupled to a harmonic-oscillator flywheel},}\
  }\href {\doibase 10.1103/PhysRevLett.123.080602} {\bibfield  {journal}
  {\bibinfo  {journal} {Phys. Rev. Lett.}\ }\textbf {\bibinfo {volume} {123}},\
  \bibinfo {pages} {080602} (\bibinfo {year} {2019})}\BibitemShut {NoStop}%
\bibitem [{\citenamefont {Kim}\ \emph {et~al.}(2022)\citenamefont {Kim},
  \citenamefont {Oh}, \citenamefont {Yang}, \citenamefont {Kim}, \citenamefont
  {Lee},\ and\ \citenamefont {An}}]{Kim.22.NP}%
  \BibitemOpen
  \bibfield  {author} {\bibinfo {author} {\bibfnamefont {J.}~\bibnamefont
  {Kim}}, \bibinfo {author} {\bibfnamefont {S.}~\bibnamefont {Oh}}, \bibinfo
  {author} {\bibfnamefont {D.}~\bibnamefont {Yang}}, \bibinfo {author}
  {\bibfnamefont {J.}~\bibnamefont {Kim}}, \bibinfo {author} {\bibfnamefont
  {M.}~\bibnamefont {Lee}}, \ and\ \bibinfo {author} {\bibfnamefont
  {K.}~\bibnamefont {An}},\ }\bibfield  {title} {\enquote {\bibinfo {title} {A
  photonic quantum engine driven by superradiance},}\ }\href
  {https://doi.org/10.1038/s41566-022-01039-2} {\bibfield  {journal} {\bibinfo
  {journal} {Nat. Photonics}\ }\textbf {\bibinfo {volume} {16}},\ \bibinfo
  {pages} {707} (\bibinfo {year} {2022})}\BibitemShut {NoStop}%
\bibitem [{\citenamefont {Bu}\ \emph {et~al.}(2023)\citenamefont {Bu},
  \citenamefont {Zhang}, \citenamefont {Ding}, \citenamefont {Li},
  \citenamefont {Zhang}, \citenamefont {Wang}, \citenamefont {Ding},
  \citenamefont {Yuan}, \citenamefont {Chen}, \citenamefont {\"Ozdemir},
  \citenamefont {Zhou}, \citenamefont {Jing},\ and\ \citenamefont
  {Feng}}]{BuJ.23.PRL}%
  \BibitemOpen
  \bibfield  {author} {\bibinfo {author} {\bibfnamefont {J.-T.}\ \bibnamefont
  {Bu}}, \bibinfo {author} {\bibfnamefont {J.-Q.}\ \bibnamefont {Zhang}},
  \bibinfo {author} {\bibfnamefont {G.-Y.}\ \bibnamefont {Ding}}, \bibinfo
  {author} {\bibfnamefont {J.-C.}\ \bibnamefont {Li}}, \bibinfo {author}
  {\bibfnamefont {J.-W.}\ \bibnamefont {Zhang}}, \bibinfo {author}
  {\bibfnamefont {B.}~\bibnamefont {Wang}}, \bibinfo {author} {\bibfnamefont
  {W.-Q.}\ \bibnamefont {Ding}}, \bibinfo {author} {\bibfnamefont {W.-F.}\
  \bibnamefont {Yuan}}, \bibinfo {author} {\bibfnamefont {L.}~\bibnamefont
  {Chen}}, \bibinfo {author} {\bibfnamefont {\ifmmode \mbox{\c{S}}\else
  \c{S}\fi{}.~K.}\ \bibnamefont {\"Ozdemir}}, \bibinfo {author} {\bibfnamefont
  {F.}~\bibnamefont {Zhou}}, \bibinfo {author} {\bibfnamefont {H.}~\bibnamefont
  {Jing}}, \ and\ \bibinfo {author} {\bibfnamefont {M.}~\bibnamefont {Feng}},\
  }\bibfield  {title} {\enquote {\bibinfo {title} {Enhancement of quantum heat
  engine by encircling a liouvillian exceptional point},}\ }\href {\doibase
  10.1103/PhysRevLett.130.110402} {\bibfield  {journal} {\bibinfo  {journal}
  {Phys. Rev. Lett.}\ }\textbf {\bibinfo {volume} {130}},\ \bibinfo {pages}
  {110402} (\bibinfo {year} {2023})}\BibitemShut {NoStop}%
\bibitem [{\citenamefont {Koch}\ \emph {et~al.}(2023)\citenamefont {Koch},
  \citenamefont {Menon}, \citenamefont {Cuestas}, \citenamefont {Barbosa},
  \citenamefont {Lutz}, \citenamefont {Fogarty}, \citenamefont {Busch},\ and\
  \citenamefont {Widera}}]{Koch.23.N}%
  \BibitemOpen
  \bibfield  {author} {\bibinfo {author} {\bibfnamefont {J.}~\bibnamefont
  {Koch}}, \bibinfo {author} {\bibfnamefont {K.}~\bibnamefont {Menon}},
  \bibinfo {author} {\bibfnamefont {E.}~\bibnamefont {Cuestas}}, \bibinfo
  {author} {\bibfnamefont {S.}~\bibnamefont {Barbosa}}, \bibinfo {author}
  {\bibfnamefont {E.}~\bibnamefont {Lutz}}, \bibinfo {author} {\bibfnamefont
  {T.}~\bibnamefont {Fogarty}}, \bibinfo {author} {\bibfnamefont
  {T.}~\bibnamefont {Busch}}, \ and\ \bibinfo {author} {\bibfnamefont
  {A.}~\bibnamefont {Widera}},\ }\bibfield  {title} {\enquote {\bibinfo {title}
  {A quantum engine in the {BEC}{\textendash}{BCS} crossover},}\ }\href
  {\doibase 10.1038/s41586-023-06469-8} {\bibfield  {journal} {\bibinfo
  {journal} {Nature}\ }\textbf {\bibinfo {volume} {621}},\ \bibinfo {pages}
  {723} (\bibinfo {year} {2023})}\BibitemShut {NoStop}%
\bibitem [{\citenamefont {Josefsson}\ \emph {et~al.}(2018)\citenamefont
  {Josefsson}, \citenamefont {Svilans}, \citenamefont {Burke}, \citenamefont
  {Hoffmann}, \citenamefont {Fahlvik}, \citenamefont {Thelander}, \citenamefont
  {Leijnse},\ and\ \citenamefont {Linke}}]{Josefsson.18.NN}%
  \BibitemOpen
  \bibfield  {author} {\bibinfo {author} {\bibfnamefont {M.}~\bibnamefont
  {Josefsson}}, \bibinfo {author} {\bibfnamefont {A.}~\bibnamefont {Svilans}},
  \bibinfo {author} {\bibfnamefont {A.~M.}\ \bibnamefont {Burke}}, \bibinfo
  {author} {\bibfnamefont {E.~A.}\ \bibnamefont {Hoffmann}}, \bibinfo {author}
  {\bibfnamefont {S.}~\bibnamefont {Fahlvik}}, \bibinfo {author} {\bibfnamefont
  {C.}~\bibnamefont {Thelander}}, \bibinfo {author} {\bibfnamefont
  {M.}~\bibnamefont {Leijnse}}, \ and\ \bibinfo {author} {\bibfnamefont
  {H.}~\bibnamefont {Linke}},\ }\bibfield  {title} {\enquote {\bibinfo {title}
  {A quantum-dot heat engine operating close to the thermodynamic efficiency
  limits},}\ }\href {https://doi.org/10.1038/s41565-018-0200-5} {\bibfield
  {journal} {\bibinfo  {journal} {Nat. Nanotechnol.}\ }\textbf {\bibinfo
  {volume} {13}},\ \bibinfo {pages} {920} (\bibinfo {year} {2018})}\BibitemShut
  {NoStop}%
\bibitem [{\citenamefont {Talkner}\ and\ \citenamefont
  {H\"anggi}(2020)}]{Talkner.20.RMP}%
  \BibitemOpen
  \bibfield  {author} {\bibinfo {author} {\bibfnamefont {P.}~\bibnamefont
  {Talkner}}\ and\ \bibinfo {author} {\bibfnamefont {P.}~\bibnamefont
  {H\"anggi}},\ }\bibfield  {title} {\enquote {\bibinfo {title} {Colloquium:
  Statistical mechanics and thermodynamics at strong coupling: Quantum and
  classical},}\ }\href {\doibase 10.1103/RevModPhys.92.041002} {\bibfield
  {journal} {\bibinfo  {journal} {Rev. Mod. Phys.}\ }\textbf {\bibinfo {volume}
  {92}},\ \bibinfo {pages} {041002} (\bibinfo {year} {2020})}\BibitemShut
  {NoStop}%
\bibitem [{\citenamefont {Longstaff}\ \emph {et~al.}(2023)\citenamefont
  {Longstaff}, \citenamefont {Jabbour},\ and\ \citenamefont
  {Brask}}]{Longstaff.23.PRA}%
  \BibitemOpen
  \bibfield  {author} {\bibinfo {author} {\bibfnamefont {B.}~\bibnamefont
  {Longstaff}}, \bibinfo {author} {\bibfnamefont {M.}~\bibnamefont {Jabbour}},
  \ and\ \bibinfo {author} {\bibfnamefont {J.}~\bibnamefont {Brask}},\
  }\bibfield  {title} {\enquote {\bibinfo {title} {Impossibility of bosonic
  autonomous entanglement engines in the weak-coupling limit},}\ }\href
  {\doibase 10.1103/PhysRevA.108.032209} {\bibfield  {journal} {\bibinfo
  {journal} {Phys. Rev. A}\ }\textbf {\bibinfo {volume} {108}},\ \bibinfo
  {pages} {032209} (\bibinfo {year} {2023})}\BibitemShut {NoStop}%
\bibitem [{\citenamefont {Gallego}\ \emph {et~al.}(2014)\citenamefont
  {Gallego}, \citenamefont {Riera},\ and\ \citenamefont
  {Eisert}}]{Gallego.14.NJP}%
  \BibitemOpen
  \bibfield  {author} {\bibinfo {author} {\bibfnamefont {R.}~\bibnamefont
  {Gallego}}, \bibinfo {author} {\bibfnamefont {A.}~\bibnamefont {Riera}}, \
  and\ \bibinfo {author} {\bibfnamefont {J.}~\bibnamefont {Eisert}},\
  }\bibfield  {title} {\enquote {\bibinfo {title} {Thermal machines beyond the
  weak coupling regime},}\ }\href {\doibase 10.1088/1367-2630/16/12/125009}
  {\bibfield  {journal} {\bibinfo  {journal} {New J. Phys.}\ }\textbf {\bibinfo
  {volume} {16}},\ \bibinfo {pages} {125009} (\bibinfo {year}
  {2014})}\BibitemShut {NoStop}%
\bibitem [{\citenamefont {Gardas}\ and\ \citenamefont
  {Deffner}(2015)}]{Gardas.15.PRE}%
  \BibitemOpen
  \bibfield  {author} {\bibinfo {author} {\bibfnamefont {B.}~\bibnamefont
  {Gardas}}\ and\ \bibinfo {author} {\bibfnamefont {S.}~\bibnamefont
  {Deffner}},\ }\bibfield  {title} {\enquote {\bibinfo {title} {Thermodynamic
  universality of quantum carnot engines},}\ }\href {\doibase
  10.1103/PhysRevE.92.042126} {\bibfield  {journal} {\bibinfo  {journal} {Phys.
  Rev. E}\ }\textbf {\bibinfo {volume} {92}},\ \bibinfo {pages} {042126}
  (\bibinfo {year} {2015})}\BibitemShut {NoStop}%
\bibitem [{\citenamefont {Kato}\ and\ \citenamefont
  {Tanimura}(2016)}]{Kato.16.JCP}%
  \BibitemOpen
  \bibfield  {author} {\bibinfo {author} {\bibfnamefont {A.}~\bibnamefont
  {Kato}}\ and\ \bibinfo {author} {\bibfnamefont {Y.}~\bibnamefont
  {Tanimura}},\ }\bibfield  {title} {\enquote {\bibinfo {title} {Quantum heat
  current under non-perturbative and non-markovian conditions: Applications to
  heat machines},}\ }\href {\doibase 10.1063/1.4971370} {\bibfield  {journal}
  {\bibinfo  {journal} {J. Chem. Phys.}\ }\textbf {\bibinfo {volume} {145}},\
  \bibinfo {pages} {224105} (\bibinfo {year} {2016})}\BibitemShut {NoStop}%
\bibitem [{\citenamefont {Strasberg}\ \emph {et~al.}(2016)\citenamefont
  {Strasberg}, \citenamefont {Schaller}, \citenamefont {Lambert},\ and\
  \citenamefont {Brandes}}]{Strasberg.16.NJP}%
  \BibitemOpen
  \bibfield  {author} {\bibinfo {author} {\bibfnamefont {P.}~\bibnamefont
  {Strasberg}}, \bibinfo {author} {\bibfnamefont {G.}~\bibnamefont {Schaller}},
  \bibinfo {author} {\bibfnamefont {N.}~\bibnamefont {Lambert}}, \ and\
  \bibinfo {author} {\bibfnamefont {T.}~\bibnamefont {Brandes}},\ }\bibfield
  {title} {\enquote {\bibinfo {title} {Nonequilibrium thermodynamics in the
  strong coupling and non-markovian regime based on a reaction coordinate
  mapping},}\ }\href {http://stacks.iop.org/1367-2630/18/i=7/a=073007}
  {\bibfield  {journal} {\bibinfo  {journal} {New J. Phys.}\ }\textbf {\bibinfo
  {volume} {18}},\ \bibinfo {pages} {073007} (\bibinfo {year}
  {2016})}\BibitemShut {NoStop}%
\bibitem [{\citenamefont {Xu}\ \emph {et~al.}(2018)\citenamefont {Xu},
  \citenamefont {Chen},\ and\ \citenamefont {Liu}}]{Xu.18.PRE}%
  \BibitemOpen
  \bibfield  {author} {\bibinfo {author} {\bibfnamefont {Y.~Y.}\ \bibnamefont
  {Xu}}, \bibinfo {author} {\bibfnamefont {B.}~\bibnamefont {Chen}}, \ and\
  \bibinfo {author} {\bibfnamefont {J.}~\bibnamefont {Liu}},\ }\bibfield
  {title} {\enquote {\bibinfo {title} {Achieving the classical carnot
  efficiency in a strongly coupled quantum heat engine},}\ }\href {\doibase
  10.1103/PhysRevE.97.022130} {\bibfield  {journal} {\bibinfo  {journal} {Phys.
  Rev. E}\ }\textbf {\bibinfo {volume} {97}},\ \bibinfo {pages} {022130}
  (\bibinfo {year} {2018})}\BibitemShut {NoStop}%
\bibitem [{\citenamefont {Restrepo}\ \emph {et~al.}(2018)\citenamefont
  {Restrepo}, \citenamefont {Cerrillo}, \citenamefont {Strasberg},\ and\
  \citenamefont {Schaller}}]{Restrepo.18.NJP}%
  \BibitemOpen
  \bibfield  {author} {\bibinfo {author} {\bibfnamefont {S.}~\bibnamefont
  {Restrepo}}, \bibinfo {author} {\bibfnamefont {J.}~\bibnamefont {Cerrillo}},
  \bibinfo {author} {\bibfnamefont {P.}~\bibnamefont {Strasberg}}, \ and\
  \bibinfo {author} {\bibfnamefont {G.}~\bibnamefont {Schaller}},\ }\bibfield
  {title} {\enquote {\bibinfo {title} {From quantum heat engines to laser
  cooling: Floquet theory beyond the born{\textendash}markov approximation},}\
  }\href {\doibase 10.1088/1367-2630/aac583} {\bibfield  {journal} {\bibinfo
  {journal} {New J. Phys.}\ }\textbf {\bibinfo {volume} {20}},\ \bibinfo
  {pages} {053063} (\bibinfo {year} {2018})}\BibitemShut {NoStop}%
\bibitem [{\citenamefont {Newman}\ \emph {et~al.}(2020)\citenamefont {Newman},
  \citenamefont {Mintert},\ and\ \citenamefont {Nazir}}]{Newman.20.PRE}%
  \BibitemOpen
  \bibfield  {author} {\bibinfo {author} {\bibfnamefont {D.}~\bibnamefont
  {Newman}}, \bibinfo {author} {\bibfnamefont {F.}~\bibnamefont {Mintert}}, \
  and\ \bibinfo {author} {\bibfnamefont {A.}~\bibnamefont {Nazir}},\ }\bibfield
   {title} {\enquote {\bibinfo {title} {Quantum limit to nonequilibrium
  heat-engine performance imposed by strong system-reservoir coupling},}\
  }\href {\doibase 10.1103/PhysRevE.101.052129} {\bibfield  {journal} {\bibinfo
   {journal} {Phys. Rev. E}\ }\textbf {\bibinfo {volume} {101}},\ \bibinfo
  {pages} {052129} (\bibinfo {year} {2020})}\BibitemShut {NoStop}%
\bibitem [{\citenamefont {Shirai}\ \emph {et~al.}(2021)\citenamefont {Shirai},
  \citenamefont {Hashimoto}, \citenamefont {Tezuka}, \citenamefont {Uchiyama},\
  and\ \citenamefont {Hatano}}]{Shirai.21.PRR}%
  \BibitemOpen
  \bibfield  {author} {\bibinfo {author} {\bibfnamefont {Y.}~\bibnamefont
  {Shirai}}, \bibinfo {author} {\bibfnamefont {K.}~\bibnamefont {Hashimoto}},
  \bibinfo {author} {\bibfnamefont {R.}~\bibnamefont {Tezuka}}, \bibinfo
  {author} {\bibfnamefont {C.}~\bibnamefont {Uchiyama}}, \ and\ \bibinfo
  {author} {\bibfnamefont {N.}~\bibnamefont {Hatano}},\ }\bibfield  {title}
  {\enquote {\bibinfo {title} {Non-markovian effect on quantum otto engine:
  Role of system-reservoir interaction},}\ }\href {\doibase
  10.1103/PhysRevResearch.3.023078} {\bibfield  {journal} {\bibinfo  {journal}
  {Phys. Rev. Research}\ }\textbf {\bibinfo {volume} {3}},\ \bibinfo {pages}
  {023078} (\bibinfo {year} {2021})}\BibitemShut {NoStop}%
\bibitem [{\citenamefont {Ivander}\ \emph {et~al.}(2022)\citenamefont
  {Ivander}, \citenamefont {Anto-Sztrikacs},\ and\ \citenamefont
  {Segal}}]{Ivander.22.PRE}%
  \BibitemOpen
  \bibfield  {author} {\bibinfo {author} {\bibfnamefont {F.}~\bibnamefont
  {Ivander}}, \bibinfo {author} {\bibfnamefont {N.}~\bibnamefont
  {Anto-Sztrikacs}}, \ and\ \bibinfo {author} {\bibfnamefont {D.}~\bibnamefont
  {Segal}},\ }\bibfield  {title} {\enquote {\bibinfo {title} {Strong
  system-bath coupling effects in quantum absorption refrigerators},}\ }\href
  {\doibase 10.1103/PhysRevE.105.034112} {\bibfield  {journal} {\bibinfo
  {journal} {Phys. Rev. E}\ }\textbf {\bibinfo {volume} {105}},\ \bibinfo
  {pages} {034112} (\bibinfo {year} {2022})}\BibitemShut {NoStop}%
\bibitem [{\citenamefont {Liu}\ and\ \citenamefont {Jung}(2022)}]{Liu.22.PRE}%
  \BibitemOpen
  \bibfield  {author} {\bibinfo {author} {\bibfnamefont {J.}~\bibnamefont
  {Liu}}\ and\ \bibinfo {author} {\bibfnamefont {K.}~\bibnamefont {Jung}},\
  }\bibfield  {title} {\enquote {\bibinfo {title} {Optimal linear cyclic
  quantum heat engines cannot benefit from strong coupling},}\ }\href {\doibase
  10.1103/PhysRevE.106.L022105} {\bibfield  {journal} {\bibinfo  {journal}
  {Phys. Rev. E}\ }\textbf {\bibinfo {volume} {106}},\ \bibinfo {pages}
  {L022105} (\bibinfo {year} {2022})}\BibitemShut {NoStop}%
\bibitem [{\citenamefont {Latune}\ \emph {et~al.}(2023)\citenamefont {Latune},
  \citenamefont {Pleasance},\ and\ \citenamefont
  {Petruccione}}]{Latune.23.PRA}%
  \BibitemOpen
  \bibfield  {author} {\bibinfo {author} {\bibfnamefont {C.}~\bibnamefont
  {Latune}}, \bibinfo {author} {\bibfnamefont {G.}~\bibnamefont {Pleasance}}, \
  and\ \bibinfo {author} {\bibfnamefont {F.}~\bibnamefont {Petruccione}},\
  }\bibfield  {title} {\enquote {\bibinfo {title} {Cyclic quantum engines
  enhanced by strong bath coupling},}\ }\href {\doibase
  10.1103/PhysRevApplied.20.024038} {\bibfield  {journal} {\bibinfo  {journal}
  {Phys. Rev. Appl.}\ }\textbf {\bibinfo {volume} {20}},\ \bibinfo {pages}
  {024038} (\bibinfo {year} {2023})}\BibitemShut {NoStop}%
\bibitem [{\citenamefont {Kaneyasu}\ and\ \citenamefont
  {Hasegawa}(2023)}]{Kaneyasu.23.PRE}%
  \BibitemOpen
  \bibfield  {author} {\bibinfo {author} {\bibfnamefont {M.}~\bibnamefont
  {Kaneyasu}}\ and\ \bibinfo {author} {\bibfnamefont {Y.}~\bibnamefont
  {Hasegawa}},\ }\bibfield  {title} {\enquote {\bibinfo {title} {Quantum otto
  cycle under strong coupling},}\ }\href {\doibase 10.1103/PhysRevE.107.044127}
  {\bibfield  {journal} {\bibinfo  {journal} {Phys. Rev. E}\ }\textbf {\bibinfo
  {volume} {107}},\ \bibinfo {pages} {044127} (\bibinfo {year}
  {2023})}\BibitemShut {NoStop}%
\bibitem [{\citenamefont {Cresser}\ and\ \citenamefont
  {Anders}(2021)}]{Cresser.21.PRL}%
  \BibitemOpen
  \bibfield  {author} {\bibinfo {author} {\bibfnamefont {J.~D.}\ \bibnamefont
  {Cresser}}\ and\ \bibinfo {author} {\bibfnamefont {J.}~\bibnamefont
  {Anders}},\ }\bibfield  {title} {\enquote {\bibinfo {title} {Weak and
  ultrastrong coupling limits of the quantum mean force gibbs state},}\ }\href
  {\doibase 10.1103/PhysRevLett.127.250601} {\bibfield  {journal} {\bibinfo
  {journal} {Phys. Rev. Lett.}\ }\textbf {\bibinfo {volume} {127}},\ \bibinfo
  {pages} {250601} (\bibinfo {year} {2021})}\BibitemShut {NoStop}%
\bibitem [{\citenamefont {Anto-Sztrikacs}\ \emph {et~al.}(2023)\citenamefont
  {Anto-Sztrikacs}, \citenamefont {Nazir},\ and\ \citenamefont
  {Segal}}]{Anto-Sztrikacs.23.PRXQ}%
  \BibitemOpen
  \bibfield  {author} {\bibinfo {author} {\bibfnamefont {N.}~\bibnamefont
  {Anto-Sztrikacs}}, \bibinfo {author} {\bibfnamefont {A.}~\bibnamefont
  {Nazir}}, \ and\ \bibinfo {author} {\bibfnamefont {D.}~\bibnamefont
  {Segal}},\ }\bibfield  {title} {\enquote {\bibinfo {title}
  {Effective-hamiltonian theory of open quantum systems at strong coupling},}\
  }\href {\doibase 10.1103/PRXQuantum.4.020307} {\bibfield  {journal} {\bibinfo
   {journal} {PRX Quantum}\ }\textbf {\bibinfo {volume} {4}},\ \bibinfo {pages}
  {020307} (\bibinfo {year} {2023})}\BibitemShut {NoStop}%
\bibitem [{\citenamefont {Weimer}\ \emph {et~al.}(2008)\citenamefont {Weimer},
  \citenamefont {Henrich}, \citenamefont {Rempp}, \citenamefont {Schr\"oder},\
  and\ \citenamefont {Mahler}}]{Weimer.08.EPL}%
  \BibitemOpen
  \bibfield  {author} {\bibinfo {author} {\bibfnamefont {H.}~\bibnamefont
  {Weimer}}, \bibinfo {author} {\bibfnamefont {M.~J.}\ \bibnamefont {Henrich}},
  \bibinfo {author} {\bibfnamefont {F.}~\bibnamefont {Rempp}}, \bibinfo
  {author} {\bibfnamefont {H.}~\bibnamefont {Schr\"oder}}, \ and\ \bibinfo
  {author} {\bibfnamefont {G.}~\bibnamefont {Mahler}},\ }\bibfield  {title}
  {\enquote {\bibinfo {title} {Local effective dynamics of quantum systems: A
  generalized approach to work and heat},}\ }\href {\doibase
  10.1209/0295-5075/83/30008} {\bibfield  {journal} {\bibinfo  {journal}
  {Europhys. Lett.}\ }\textbf {\bibinfo {volume} {83}},\ \bibinfo {pages}
  {30008} (\bibinfo {year} {2008})}\BibitemShut {NoStop}%
\bibitem [{\citenamefont {Alipour}\ \emph {et~al.}(2022)\citenamefont
  {Alipour}, \citenamefont {Rezakhani}, \citenamefont {Chenu}, \citenamefont
  {del Campo},\ and\ \citenamefont {Ala-Nissila}}]{Alipour.22.PRA}%
  \BibitemOpen
  \bibfield  {author} {\bibinfo {author} {\bibfnamefont {S.}~\bibnamefont
  {Alipour}}, \bibinfo {author} {\bibfnamefont {A.~T.}\ \bibnamefont
  {Rezakhani}}, \bibinfo {author} {\bibfnamefont {A.}~\bibnamefont {Chenu}},
  \bibinfo {author} {\bibfnamefont {A.}~\bibnamefont {del Campo}}, \ and\
  \bibinfo {author} {\bibfnamefont {T.}~\bibnamefont {Ala-Nissila}},\
  }\bibfield  {title} {\enquote {\bibinfo {title} {Entropy-based formulation of
  thermodynamics in arbitrary quantum evolution},}\ }\href {\doibase
  10.1103/PhysRevA.105.L040201} {\bibfield  {journal} {\bibinfo  {journal}
  {Phys. Rev. A}\ }\textbf {\bibinfo {volume} {105}},\ \bibinfo {pages}
  {L040201} (\bibinfo {year} {2022})}\BibitemShut {NoStop}%
\bibitem [{\citenamefont {Strasberg}\ and\ \citenamefont
  {Winter}(2021)}]{Strasberg.21.PRXQ}%
  \BibitemOpen
  \bibfield  {author} {\bibinfo {author} {\bibfnamefont {P.}~\bibnamefont
  {Strasberg}}\ and\ \bibinfo {author} {\bibfnamefont {A.}~\bibnamefont
  {Winter}},\ }\bibfield  {title} {\enquote {\bibinfo {title} {First and second
  law of quantum thermodynamics: A consistent derivation based on a microscopic
  definition of entropy},}\ }\href {\doibase 10.1103/PRXQuantum.2.030202}
  {\bibfield  {journal} {\bibinfo  {journal} {PRX Quantum}\ }\textbf {\bibinfo
  {volume} {2}},\ \bibinfo {pages} {030202} (\bibinfo {year}
  {2021})}\BibitemShut {NoStop}%
\bibitem [{\citenamefont {Parrondo}\ \emph {et~al.}(2015)\citenamefont
  {Parrondo}, \citenamefont {Horowitz},\ and\ \citenamefont
  {Sagawa}}]{Parrondo.15.NP}%
  \BibitemOpen
  \bibfield  {author} {\bibinfo {author} {\bibfnamefont {J.}~\bibnamefont
  {Parrondo}}, \bibinfo {author} {\bibfnamefont {J.}~\bibnamefont {Horowitz}},
  \ and\ \bibinfo {author} {\bibfnamefont {T.}~\bibnamefont {Sagawa}},\
  }\bibfield  {title} {\enquote {\bibinfo {title} {Thermodynamics of
  information},}\ }\href {http://dx.doi.org/10.1038/nphys3230} {\bibfield
  {journal} {\bibinfo  {journal} {Nat. Phys.}\ }\textbf {\bibinfo {volume}
  {11}},\ \bibinfo {pages} {131} (\bibinfo {year} {2015})}\BibitemShut
  {NoStop}%
\bibitem [{\citenamefont {Deffner}\ and\ \citenamefont
  {Lutz}()}]{Deffner.12.A}%
  \BibitemOpen
  \bibfield  {author} {\bibinfo {author} {\bibfnamefont {S.}~\bibnamefont
  {Deffner}}\ and\ \bibinfo {author} {\bibfnamefont {E.}~\bibnamefont {Lutz}},\
  }\href {https://arxiv.org/abs/1201.3888} {\bibinfo  {journal}
  {arXiv:1201.3888}\ }\BibitemShut {NoStop}%
\bibitem [{\citenamefont {Deffner}\ and\ \citenamefont
  {Jarzynski}(2013)}]{Deffner.13.PRX}%
  \BibitemOpen
\bibfield  {journal} {  }\bibfield  {author} {\bibinfo {author} {\bibfnamefont
  {S.}~\bibnamefont {Deffner}}\ and\ \bibinfo {author} {\bibfnamefont
  {C.}~\bibnamefont {Jarzynski}},\ }\bibfield  {title} {\enquote {\bibinfo
  {title} {Information processing and the second law of thermodynamics: An
  inclusive, hamiltonian approach},}\ }\href {\doibase
  10.1103/PhysRevX.3.041003} {\bibfield  {journal} {\bibinfo  {journal} {Phys.
  Rev. X}\ }\textbf {\bibinfo {volume} {3}},\ \bibinfo {pages} {041003}
  (\bibinfo {year} {2013})}\BibitemShut {NoStop}%
\bibitem [{\citenamefont {Esposito}\ \emph {et~al.}(2010)\citenamefont
  {Esposito}, \citenamefont {Lindenberg},\ and\ \citenamefont {den
  Broeck}}]{Esposito.10.NJP}%
  \BibitemOpen
  \bibfield  {author} {\bibinfo {author} {\bibfnamefont {M.}~\bibnamefont
  {Esposito}}, \bibinfo {author} {\bibfnamefont {K.}~\bibnamefont
  {Lindenberg}}, \ and\ \bibinfo {author} {\bibfnamefont {C.~Van}\ \bibnamefont
  {den Broeck}},\ }\bibfield  {title} {\enquote {\bibinfo {title} {Entropy
  production as correlation between system and reservoir},}\ }\href {\doibase
  10.1088/1367-2630/12/1/013013} {\bibfield  {journal} {\bibinfo  {journal}
  {New J. Phys.}\ }\textbf {\bibinfo {volume} {12}},\ \bibinfo {pages} {013013}
  (\bibinfo {year} {2010})}\BibitemShut {NoStop}%
\bibitem [{\citenamefont {Van~den Broeck}(2005)}]{Broeck.05.PRL}%
  \BibitemOpen
  \bibfield  {author} {\bibinfo {author} {\bibfnamefont {C.}~\bibnamefont
  {Van~den Broeck}},\ }\bibfield  {title} {\enquote {\bibinfo {title}
  {Thermodynamic efficiency at maximum power},}\ }\href {\doibase
  10.1103/PhysRevLett.95.190602} {\bibfield  {journal} {\bibinfo  {journal}
  {Phys. Rev. Lett.}\ }\textbf {\bibinfo {volume} {95}},\ \bibinfo {pages}
  {190602} (\bibinfo {year} {2005})}\BibitemShut {NoStop}%
\bibitem [{\citenamefont {Allahverdyan}\ \emph {et~al.}(2004)\citenamefont
  {Allahverdyan}, \citenamefont {Balian},\ and\ \citenamefont
  {Nieuwenhuizen}}]{Allahverdyan.04.EPL}%
  \BibitemOpen
  \bibfield  {author} {\bibinfo {author} {\bibfnamefont {A.~E.}\ \bibnamefont
  {Allahverdyan}}, \bibinfo {author} {\bibfnamefont {R.}~\bibnamefont
  {Balian}}, \ and\ \bibinfo {author} {\bibfnamefont {Th.~M}\ \bibnamefont
  {Nieuwenhuizen}},\ }\bibfield  {title} {\enquote {\bibinfo {title} {Maximal
  work extraction from finite quantum systems},}\ }\href {\doibase
  10.1209/epl/i2004-10101-2} {\bibfield  {journal} {\bibinfo  {journal}
  {Europhys. Lett.}\ }\textbf {\bibinfo {volume} {67}},\ \bibinfo {pages}
  {565--571} (\bibinfo {year} {2004})}\BibitemShut {NoStop}%
\bibitem [{\citenamefont {Tirone}\ \emph {et~al.}()\citenamefont {Tirone},
  \citenamefont {Salvia}, \citenamefont {Chessa},\ and\ \citenamefont
  {Giovannetti}}]{Tirone.22.A}%
  \BibitemOpen
  \bibfield  {author} {\bibinfo {author} {\bibfnamefont {S.}~\bibnamefont
  {Tirone}}, \bibinfo {author} {\bibfnamefont {R.}~\bibnamefont {Salvia}},
  \bibinfo {author} {\bibfnamefont {S.}~\bibnamefont {Chessa}}, \ and\ \bibinfo
  {author} {\bibfnamefont {V.}~\bibnamefont {Giovannetti}},\ }\href
  {https://arxiv.org/abs/2211.02685} {\bibinfo  {journal} {arXiv:2211.02685}\
  }\BibitemShut {NoStop}%
\end{thebibliography}
%

\end{document}